\documentclass[a4paper,11pt]{article}
\pdfoutput=1 

\usepackage{jinstpub} 
\usepackage{graphicx}
\usepackage{newtxtext,newtxmath}

\title{\boldmath Fast simulation of avalanche and streamer in GEM detector using hydrodynamic approach}

\author[a,b,1]{Prasant Kumar Rout,\note{Corresponding author.}}
\author[a,b]{Jaydeep Datta,}
\author[a,b]{Promita Roy,}
\author[c]{Purba Bhattacharya,}
\author[a,b]{Supratik Mukhopadhyay,}
\author[a,b]{Nayana Majumdar,}
\author[a,b]{Sandip Sarkar}
\affiliation[a]{Saha Institute Of Nuclear Physics, \\1/AF Saltlake Kolkata 700064, India}
\affiliation[b]{Homi Bhaba National Institute, Training School Complex, Anushaktinagar, Mumbai 400094, India}
\affiliation[c]{Department of Physics, University of Calcutta, Rajabazar, Kolkata, West Bengal 700009, India}

\emailAdd{prasant.rout@saha.ac.in}

\abstract{A fast, hydrodynamic numerical model has been developed on the COMSOL Multiphysics platform to simulate the evolution and dynamics of charged particles in gaseous ionization detectors based on the Gaseous Electron Multipliers (GEM).
Effects of using two-dimensional (2D), 2D axisymmetric and three-dimensional(3D) models of the detectors have been analyzed to choose the optimum configuration.
The chosen model has been used to follow the entire operating regime of single, double and triple GEM detectors, including  avalanche and streamer mode operations.
The accumulation of space charge, its contribution towards the distortion of the applied electric field and production of streamers have been investigated in fair detail using the optimized model.}

\keywords{Detector modelling and simulations II, Micropattern gaseous detectors (GEM), Electron multipliers (gas), Gaseous detectors, Charge transport and multiplication in gas, Avalanche-induced secondary effects}

\arxivnumber{2011.02988} 

\begin{document}
\maketitle
\flushbottom
\section{Introduction}
\label{sec:Intro}
In laboratories around the world, gaseous ionization detectors are used in nuclear and particle physics experiments for their good position ($\sim$30 $\mu m$) and time resolution in the nanosecond range  \cite{Titov:2007fm}.
The Micro-Pattern Gaseous Detectors (MPGD) are well known for their excellent tracking \cite{George:2015vfj} and triggering \cite{Alfonsi:2004jm} capabilities in high rate experiments.
The Gas Electron Multiplier (GEM) \cite{Sauli1997} detector belongs to the family of MPGDs \cite{Hoch2004}.
These detectors have been proven to have high rate handling capability and radiation hardness and, thus, are being used in the high particle$-$flux environments seen in COMPASS \cite{Abbon}, LHCb \cite{LHCb2001}, ALICE \cite{ALICE1999} and CMS \cite{CMS2015} experiments at CERN.
The improvement of readout technology and related electronics so far allows users to achieve fast response, excellent spatial ( $\sim$110 $\mu m$) and time resolution (4 ns) \cite{Calabria:2016lez} through the use of these detectors.

Despite their successes in a wide variety of experiments, optimum operation of GEM detectors necessitates careful and detailed study on various aspects of the detector.
The parameters of the GEM foil, overall geometry of the detector, the gas mixture to be used etc., are determined by the experimental requirements of tracking and timing precisions.
While the ultimate choice of design parameters can be finalized only after successful experimental demonstration, a suitable numerical simulation tool can aid the decision process significantly.
Usually, the Garfield++ simulation framework \cite{Garfield} is used for the latter purpose.
This framework incorporates Heed \cite{Heed} for estimating primary ionization and MAGBOLTZ \cite{Magboltz} to analyze the effects of using different gas mixtures.
One among several finite element (such as \cite{Comsol}) or boundary element codes (such as \cite{neBEM}) is used to estimate the electric field.
The resulting simulation framework is extremely capable and is used by a large number of scientists in the community.

Garfield++ adopts a Lagrangian description to simulate the evolution of charged particles within gaseous ionization detectors and resulting signal generation.
It follows the development of various important processes, starting from the creation of the primaries in the drift volume.
Transport, gain and loss of the charged particles are followed at different levels of detail to simulate various processes and estimate figures of merit of the detector such as efficiency, gain, transmission, ion backflow etc \cite{Bhattacharya:2016vni, Bhattacharya:2017yaj}.
Recently, attempts have been made to incorporate complex issues such as charging up with moderate success \cite{Correia:2014vla}.
The Lagrangian approach, while being the most realistic, requires significant computational resources, in terms of both hardware and time.
As a result, problems involving space charge effects, development of streamers / discharges etc. are difficult to address using this simulation paradigm.
Hence, a relatively less detailed model based on the Eulerian description, has been adopted in this work to simulate the response of a detector in the entire parameter space in which it is likely to be used.
The advantage of using such a model is especially evident in situations where the particle model suffers the most - when the number of charged particles is large in number, close to the formation of streamers.
Due to the nature of the mathematical model, it has also been possible to incorporate effects due to space charge effortlessly throughout the evolution of the phenomena.
It is well known that the slow movement of ions influences subsequent development of space charge in the active volume of the detector that distorts the applied electric field and subsequent movement of charges.
According to the simulations presented in this work, positive streamers form due to uncontrolled charge multiplication as a result of the accumulation of space charges inside the GEM holes.

While particle simulation of avalanches in GEM by electrons and ions have been very well studied using the Garfield++ simulation framework, as far as we know, the development of streamers including the effects of space charge has not been performed by it yet for the reasons mentioned above.
It seems, in contrast, that the present approach can work as a relatively fast method to investigate the behavior of gaseous ionization detectors under complex situations.
While the approach is not expected to yield very detailed description of the various processes taking place within the device, the merit of the proposed approach is the ability to capture the essence of the complex developments in a reasonably short time.
The simulation framework has been built on the platform of a commercial Finite Element Method (FEM) package, COMSOL Multiphysics \cite{Comsol} and can be considered as an extension of the idea proposed by P. Fonte \cite{Fonte_RPC}.
Among other possibilities explored by Fonte, the use of a simple 2D axisymmetric hydrodynamic model was proposed \cite{avalanches_streamer} for simulation of dynamics of charged fluids in GEM detectors. 
This framework has been further elaborated and extended recently \cite{Fillipo} to simulate complex device dynamics of a number of MPGDs, including GEMs. 
In the present work, various possibilities have been explored and compared to arrive at a representation of GEM-based detectors that is both realistic and computationally inexpensive.

The complete simulation model proposed here utilizes the primary ionization produced by \cite{Heed} and electron transport parameters in the gas mixtures produced by \cite{Magboltz}.
The present study includes the results of using an Ar + CO$_{2}$ gas mixture in proportions $70:30$ by volume.
Attempts have been made to understand the advantages and limitations of the
proposed model from mathematical and numerical points of view.
The mathematical model adapted for these studies has been described in section \ref{sec:MathModel}.
Optimization of the numerical
model has been presented in section \ref{sec:NumImp}.
Section \ref{sec:ResDis} contains the results of the studies related to different operating regimes and analysis of streamer formation.
Attempts are made in this section to illustrate the importance of the effect of space charge at various phases of detector operation, especially in the formation of streamers.
Finally, concluding remarks are presented in section \ref{sec:Conc}.
\section{Mathematical model}
\label{sec:MathModel}
The mathematical model representing the evolution of different processes inside a GEM detector have been considered to be, broadly speaking, that of mass transport.
For the present work, the movement of negative and positively charged particles are considered to be analogous to mass transport of dissolved solute species in a background solvent gas mixture.
Mass transport is known to take place as manifestation of various laws of conservation under the action of driving forces such as chemical, convective, diffusive, electrical, frictional etc.
In the present work, all of the forces mentioned above are likely to have some role to play in one form or the other, determining the dynamics of the system under consideration.
Interestingly, the applied electric field, which has a considerable influence on the system dynamics, in its turn is modified by the charged particles themselves.
The attempt here is to model each of these non-linearly coupled components to an accuracy so that the response of the GEM detector may be predicted with reasonable confidence.

The hydrodynamics of the overall system is described by the drift-diffusion equations.
In particular, the transport of charged species has been modeled by using the Transport of Dilute Species (TDS) module of COMSOL.
The module assumes the species (electron or ion) concentration to be much smaller than that of the solvent (neutral gas molecules).
The  following  drift$-$diffusion$-$reaction equations \ref{eq:BTE1} $-$  \ref{eq:BTE4} are solved in COMSOL for the field dependent charge transportation.
\begin{equation}
\label{eq:BTE1}
\centering
{ \frac{\partial{c_{i}}}{\partial{t}} + {\vec{\nabla}}\cdot({-D_{i}\vec{\nabla}{c_{i}} + \vec{u_{i}}}{c_{i}}) = R_{i}}
\end{equation}
\begin{equation}
\label{eq:BTE2}
\centering
{{R_{i}} = {S_{e}} + {S_{ph}}}
\end{equation}
\begin{equation}
\label{eq:BTE3}
\centering
 {{S_{e}} = ({\alpha{(\vec{E})}} - {\eta{(\vec{E})}}){\mid{\vec{u_{e}}}}\mid{n_{e}}}
\end{equation}
\begin{equation}
\label{eq:BTE4}
\centering
S_{ph} = \xi QE_{gas}\mu\Psi_{0}
\end{equation}
where $c_{i}$, $R_{i}$, $\vec{u_{i}}$ and $D_{i}$ represents the concentration, rate of production, drift velocity and diffusion coefficients for charged species respectively.
In our model we have considered electrons and ionized gas molecules as charged species.
$S_{e}$ is the source term for electrons due to Townsend ionization, while $S_{ph}$ represent the source due to photo-ionization.
In the equation \ref{eq:BTE3}, $n_e$, $\vec{u_{e}}$, $\alpha$ and $\eta$ represents concentration, drift velocity, the first Townsend and attachment coefficients of electron respectively.
Photons emitted due to the de-excitation of the gas mixture component with higher ionization energy, ionizes the other gases of the mixture with low ionization energy.
The terms $\mu$ and ${\Psi_{0}}$ correspond to the photo absorption coefficient and the UV photon flux generated in the detector volume.
$QE_{gas}$ gives quantum efficiency of the gas mixture to produce electrons due to photon absorption.
$\xi$ is the fraction of the excited states which can ionize the gas. 

The transport coefficients for electrons are calculated as a function of electric field from MAGBOLTZ for Ar + CO$_{2}$  gas  mixtures and are included in the simulation framework.
In 2D axisymmetric model, the diagonal elements of the diffusion tensor gives the radial (along r-coordinate) and axial (along z-coordinate) diffusion coefficients.
In 3D model, the diagonal elements represent X, Y and Z components of diffusion.
The off-diagonal elements are taken as zero due to the absence of magnetic field in the volume considered.

In Ar + CO$_{2}$ gas mixtures, CO$_2$ plays the role of photon quencher.
These photons are emitted while de-excitation of excited Ar or CO$_2$ molecules.
As the ionization energy of Ar is higher than that of CO$_2$, a fraction of the absorbed photons by CO$_2$, emitted while de-excitation of the excited Ar atoms, ionizes the CO$_{2}$ molecules. 
This process happens alongside Townsend ionization.

Capeill{\`{e}}re et al. has shown in  \cite{Capeillere2008}, that photon transport in the gas volume can be described using the following equation
\begin{equation}
\label{eq:RTE}
\centering
{\vec{\nabla}(-c{\vec{\nabla}}{\Psi_{0}}) + a{\Psi_{0}} = f}
\end{equation}
where $c$ = $\frac{1}{3\mu_{all}}$, \hskip 2pt $f$ = ${\delta}$ $S_{e}$, \hskip 2pt $a$ = $\mu_{all}$.   
${\delta}$ is the number of excited neutral Ar atoms per ionized Ar molecule.
The average photo-absorption coefficient($\mu_{all}$) of CO$_{2}$ has been calculated considering all the excited states of Ar using the photo absorption cross section from \cite{Sahin:2014haa}.
Equation \ref{eq:RTE} represents the diffusion-like approximation to the photon flux(${\Psi_{0}}$) field.
In the present simulation model, the photon propagation is given by the "Coefficient Form Partial Differential Equation" module of COMSOL.

As all the photons absorbed by CO$_2$ will not be able to ionize, we have calculated $\mu$ considering only those excited states of Ar which has energy more than the ionization energy of CO$_2$.
We have also calculated the fraction of these excited states of Ar from the data \cite{NIST_Ar} and used it as a multiplicative factor $\xi$, in $S_{ph}$ as shown in equation \ref{eq:BTE4}.
In table \ref{tab:table1} we have summarized all the parameters that have been used regarding photo-ionization.
\begin{table}[h!]
	\begin{center}
		\begin{tabular}{|c|c|}
			\hline
			\textbf{Parameter} & \textbf{Value} \\
			\hline
			 $\delta$ & 10$^{-4}$ \\
			 \hline
			 QE$_{\rm gas}$ & 10$^{-4}$  \\
			 \hline
			 $\mu_{\rm all}$ & 981/cm  \\
			\hline
			$\mu$ & 831/cm \\
			\hline
			$\xi$ & 0.133 \\
			\hline
		\end{tabular}
		\caption{Parameters used for photo-ionization and photon transport}
		\label{tab:table1}
	\end{center}
\end{table} 

Open boundary conditions have been implemented on many of the boundaries for the charge transport of electrons and ions within the model volume.
It is based on the assumption that, the absorption of electrons takes place in the Cu surface of GEM foils and the anode by drifting through it without any resistance.
In a similar fashion, absorption of ions also takes place in the Cu surface of the GEM foils and cathode during their movement.
The boundaries of the simulation volume are also made open so that the charged fluid flows out freely on reaching them.
In addition, the dielectric material polyimide has been assumed to be non-porous and a "No Flux" boundary condition has been implemented on all polyimide surfaces. This implies that there is no mass flow across the dielectric material. However, the fluid can touch and flow along the surface of the dielectric.

The electric field at each time step is computed by the "Electrostatics" module of COMSOL.
The applied electric field gets distorted due to the presence of space charges in the gas volume and, as a result, has been computed at each time step.
The equations for calculating the electric potential(V) and field$(\vec{E})$ are given as follows. 
\begin{equation}
\label{eq:Field}
\centering
  {{\vec{E}} = -\vec{\nabla}{V}}
\end{equation}
\begin{equation}
\label{eq:poisson}
\centering
{{\vec{\nabla}}.D = {\rho_{v}}} 
\end{equation}
where $\vec{D}$ is the electric displacement vector, 
$\vec{D} = \epsilon_{0} \epsilon_{r} \vec{E}$, $\epsilon_{0}$ is the permittivity of vacuum or air and $\epsilon_{r}$ is the relative permittivity of the material used in the detector geometry.
$\rho_{v}$ is the volume space charge density in the gas volume and is calculated as, $\rho_{v}$ = $\frac{Q_{e}}{\epsilon_{0}}(n_{i} - n_{e}) $, $Q_{e}$ being the charge of an electron and $n_i$ represents the concentration of ion.
\section{Numerical Implementation}
\label{sec:NumImp}
Since the GEM hole has natural axisymmetry, adoption of a 2D axisymmetric geometry is expected to yield reasonable results.
However, a detailed study with several possible geometries, namely 2D, 2D axisymmetric and 3D, was carried out to establish the advantages and short-comings of each.
\subsection{Optimization of the model geometry}
\label{subsec:Geom}
\begin{figure}[htbp]
    \centering
    \includegraphics[width=\linewidth]{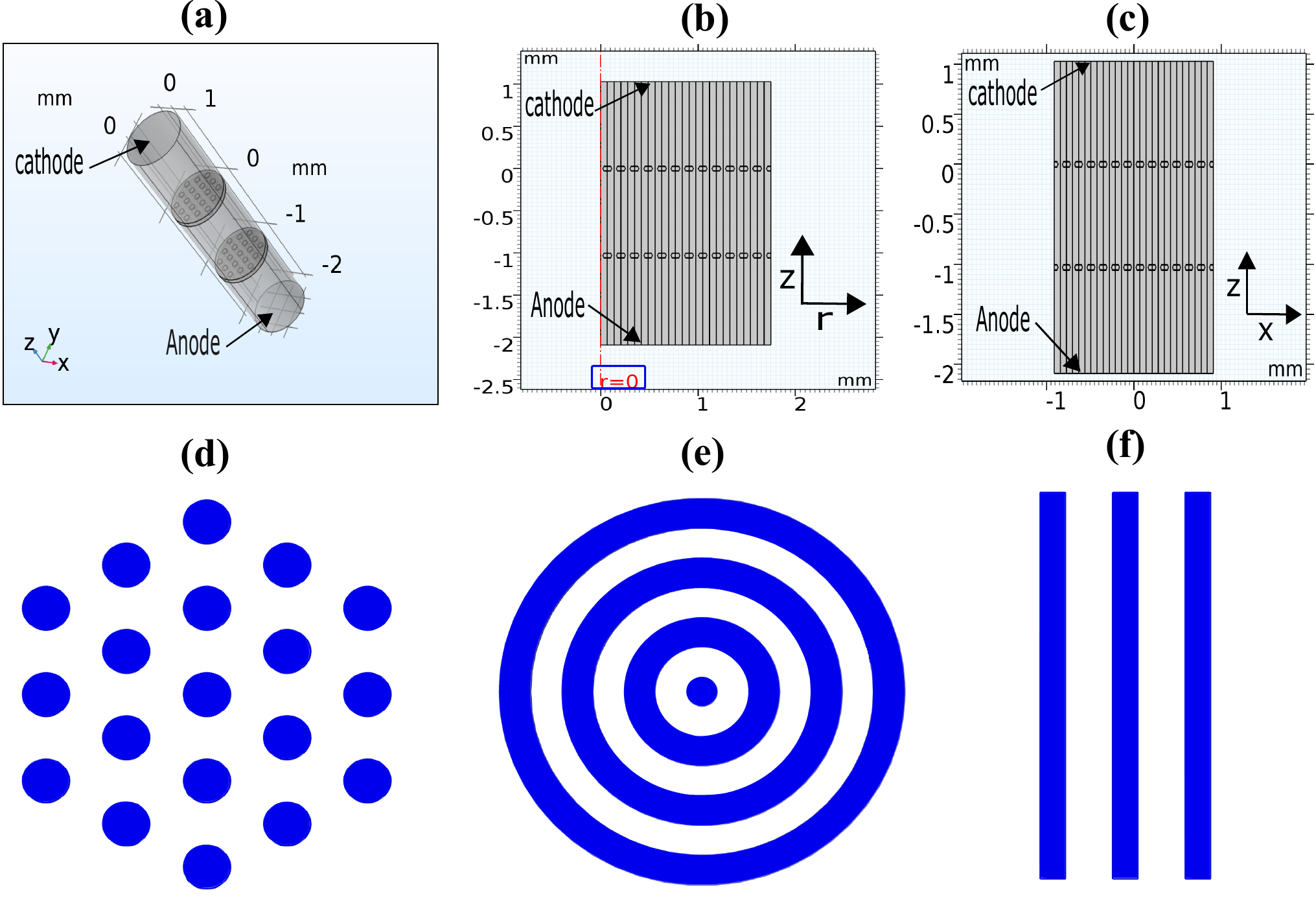}
    \caption{(a) 3D model, (b) 2D axisymmetry model and (c) 2D model, of a double GEM. (d) holes in 3D model, (e) central hole and circular channels in 2D axisymmetry model and (f) straight channels in 2D model. The blue regions in (d), (e) and (f) represent the gas volume and are the schematic representations of the real geometries in (a), (b) and (c) respectively. }
    \label{fig:fig1}
\end{figure}

The 3D model is, by its nature, the most complete description of problems being addressed in this paper.
3D models of single and multi GEM configurations have been developed to study the growth of avalanche and streamers.
A representative 3D model of a double GEM device has been shown in figure \ref{fig:fig1}(a).
In this particular model, as in all the other computations discussed in this section, the standard bi-conical hole GEM was used.
Thus, 5 $\mu m$ copper on both sides of a 50 $\mu m$ polyimide has been assumed.
The model has been made with a large number of holes having a pitch of 140 $\mu m$, so that the charged particles are not lost by the sides of the device.
This extended model also allows us to study the way the charges are being shared by different holes of the foils.

In addition to the 3D models, 2D and 2D axisymmetric models of single, double and triple GEM configurations have been developed and studied.
As examples, double GEM configurations of 2D axisymmetric and 2D models have been illustrated in figures \ref{fig:fig1}(b) and (c), respectively.
In the following sections comparison among these different possible models is presented.
\subsection{Electric potential and field}
\label{subsec:Electric Field}
For the detector geometries described in section \ref{subsec:Geom}, the electric potential and field is calculated by the FEM solver in the gas volume by utilizing extremely fine meshing of the model geometries.
Copper (Cu) and Kapton materials have been used in the GEM foils to compute the electric field.
Table \ref{tab:table2} shows the potential, field and gas gap configurations of single, double and triple GEM structures.
\begin{table}[h!]
  \begin{center}
    \begin{tabular}{|c|c|c|c|c|c|}
    \hline
      \textbf{GEM Structures} & \textbf{Range} & \textbf{Drift Field} & \textbf{Transfer  Field} & \textbf{Induction Field} & \textbf{Gap} \\
      & $\Delta{V}_{\rm GEM}$(V) & (kV/cm) & (kV/cm) & (kV/cm) & (mm)\\
      \hline
      Single GEM & 450 - 540 & 2 & -   & 3.5 & 1:1\\
      \hline
      Double GEM & 360 - 450 & 2 & 3.5 & 3.5 & 1:1:1\\
      \hline
      Triple GEM & 310 - 400 & 2 & 3.5 & 3.5 & 1:1:1:1\\
      \hline
    \end{tabular}
    \caption{Electric potential, field and gas gap configurations for single, double and Triple GEM structures}
    \label{tab:table2}
  \end{center}
\end{table}

\begin{figure}[htbp]
\centering
\includegraphics[width=0.49\linewidth]{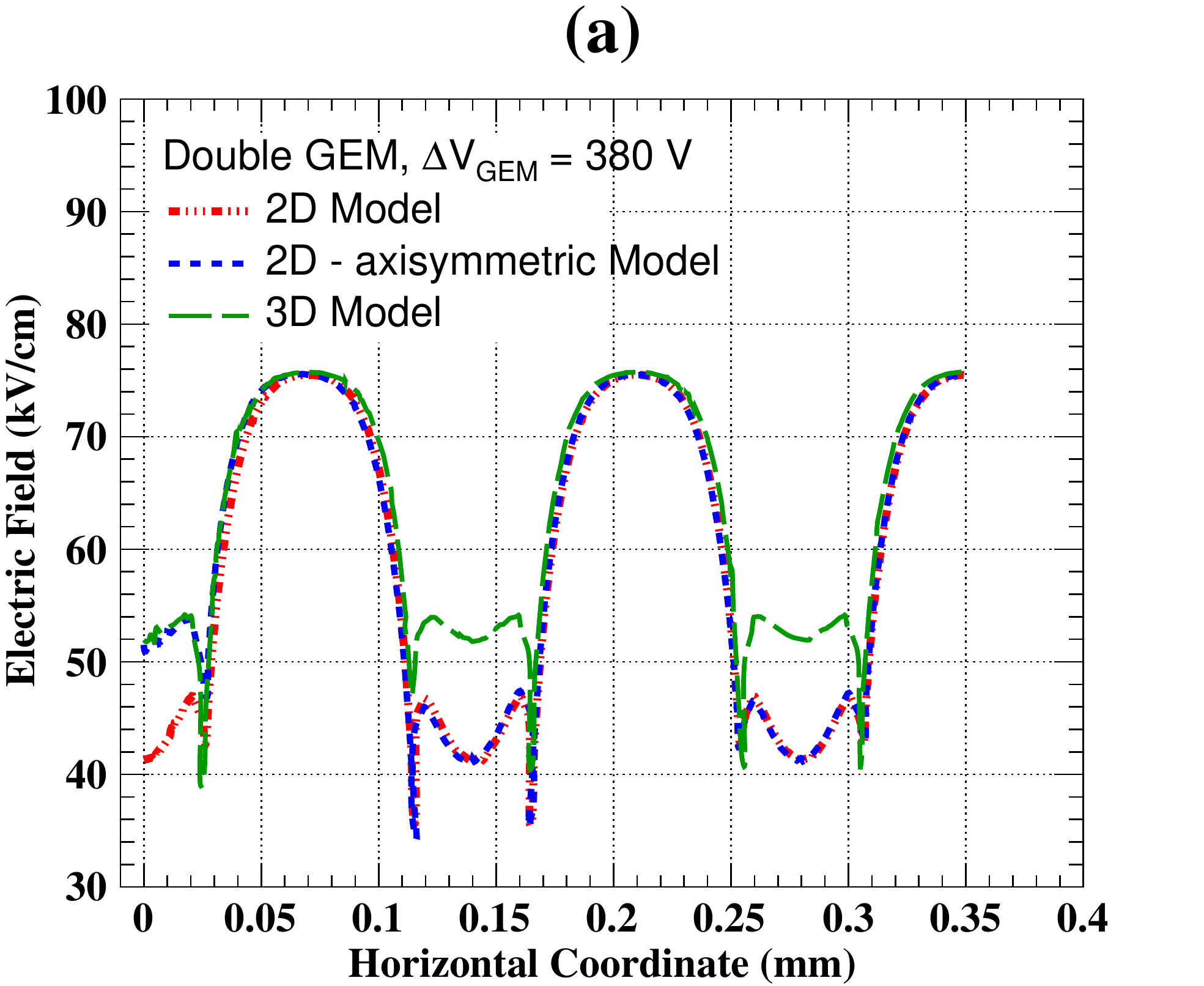}
\includegraphics[width=0.49\linewidth]{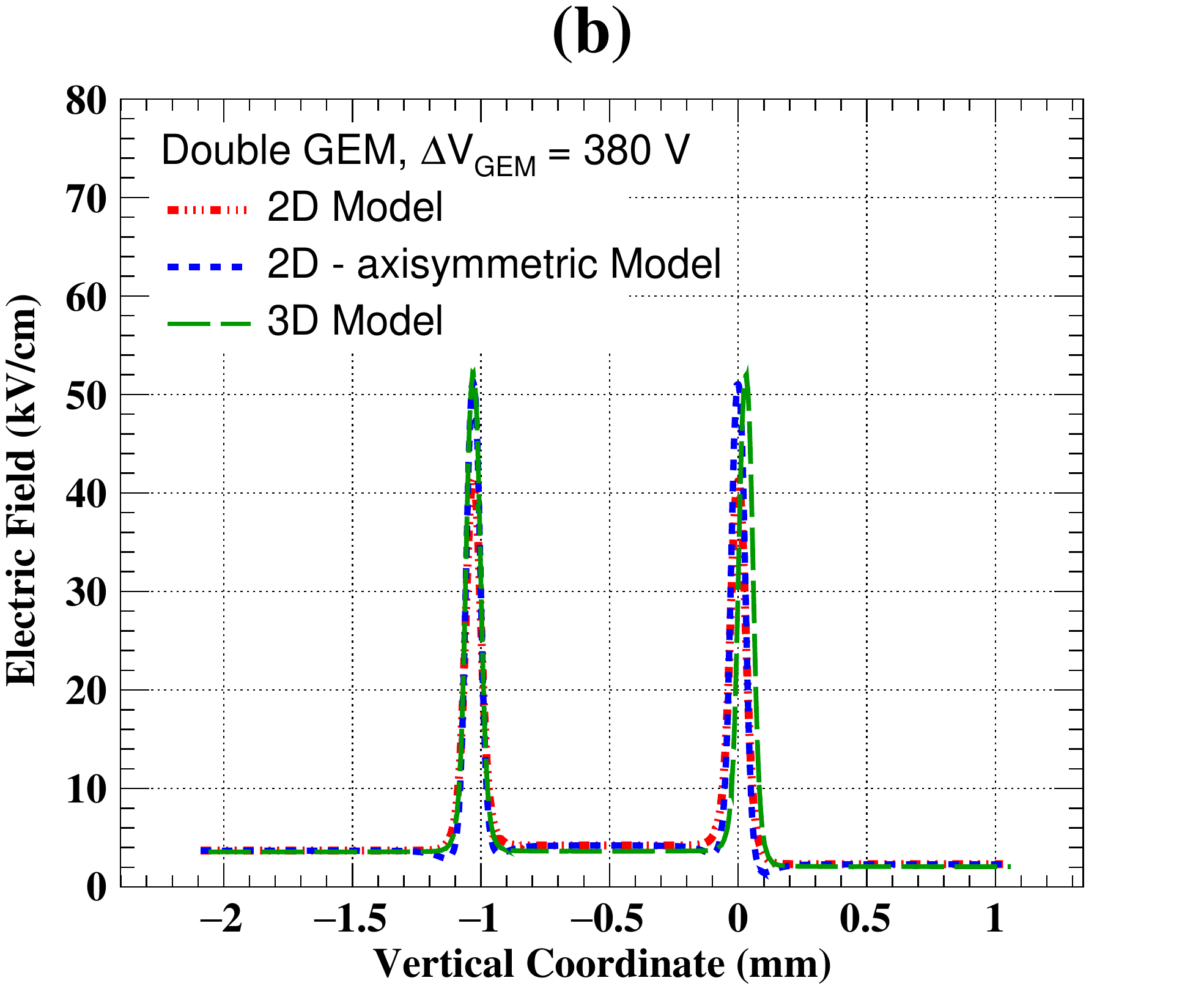}
\caption{(a) Total electric field along the middle of the dielectric in a double GEM using different models. (b) Total electric field along hole axis in a double GEM using different models}
\label{fig:fig2}
\end{figure}

Figures \ref{fig:fig2}(a) and (b), show the comparison of electric fields computed using COMSOL for a double GEM ($\Delta{V}_{\rm GEM}$ = 380V) for 3D, 2D axisymmetric and 2D models. It is interesting to note that the central hole of the 2D axisymmetric model as shown in the figure \ref{fig:fig1}(e) has the same field distribution as that of the holes of 3D model in figure \ref{fig:fig1}(d).
However, the entire 2D model and off-centre holes of the 2D axisymmetric model differ significantly from their 3D counterpart.
The deviation is expected because, due to the nature of the model, all 2D `holes' and all non-central `holes' of the axisymmetric models are in fact channels instead of holes as shown in the figures \ref{fig:fig1}(e) and (f) respectively.
The maximum difference in field is around 20\% for the case considered in the plots.
This difference in field values among different models have very significant consequences in the dynamics of charge density evolution because all the transport parameters are strongly dependent on the field values.
As a result, the estimates of the detector properties of our interest are expected to vary significantly for different models.
However, since the holes of 3D model and the central hole of axisymmetric model have similar field strengths, and additionally, the overall dynamics depends more on what happens at the central hole, the 2D axisymmetric model can be a reasonable alternative to the full-fledged 3D model. Moreover, using few simplifying assumptions, it has also been possible to scale the axisymmetric estimates of
electron multiplication so that they are comparable to the 3D estimates. The details of obtaining such scaling factors have been provided in Appendix \ref{sec:Appendix 1}.
\subsection{Validity of assumption of axisymmetry}
\begin{figure}[htbp]
\centering
\includegraphics[width=0.50\linewidth]{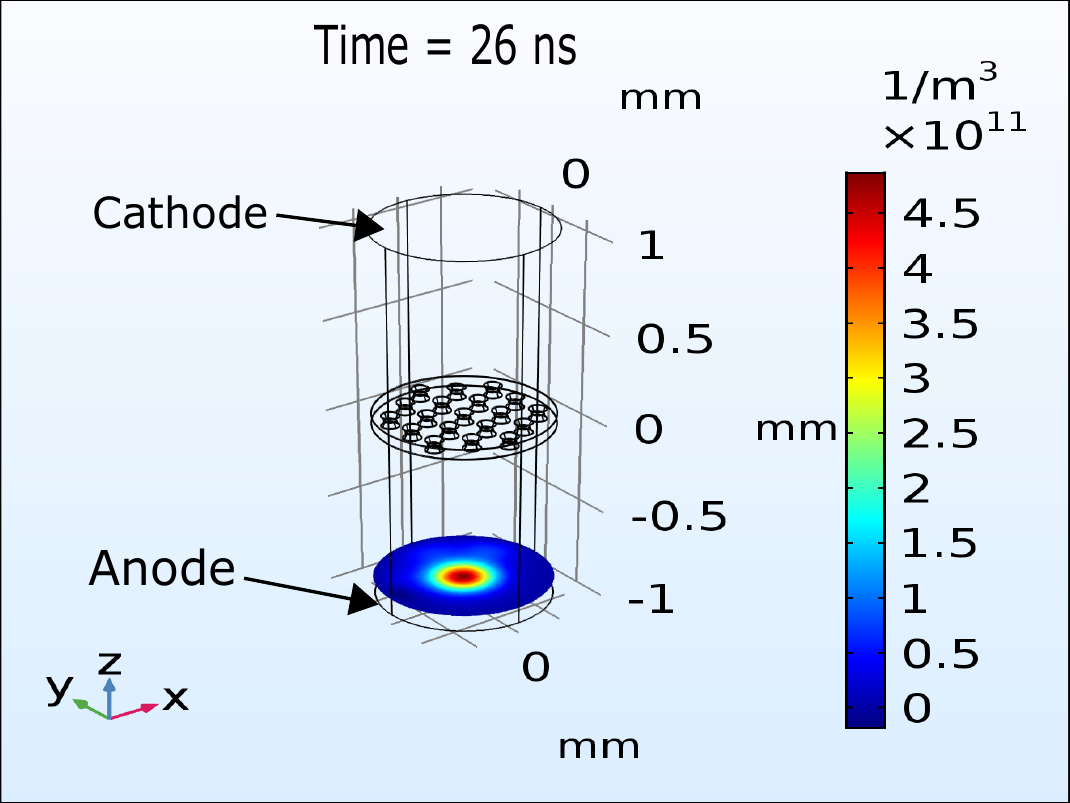}
\caption{Axisymmetric nature of the shape of avalanche in a single GEM 3D model close to the readout anode.}
\label{fig:fig3}
\end{figure}

In order to confirm that the assumption of axisymmetry is reasonable throughout the evolution of charge transport in such devices, several test executions have been performed using the 3D models. For example, the evolution of avalanche using the 3D model, as shown
in figure \ref{fig:fig3}, clearly shows the axisymmetric nature of the avalanche even when it is very
close to the readout anode. Although this axial symmetry may be disturbed during the formation of streamer, for usual operating conditions, the rotational symmetry is unlikely to be lost.
\subsection{Optimum model and necessary computational resources}
The aim of the numerical experiments described above was to identify a numerical model that would be economic, fast (simulation time not more than an hour on a present day desktop) and which retains most of the important physics processes occurring in a GEM detector.
The 3D models, although the most accurate representations, were out of contention because of the computational resources they demand (several tens of GBs of RAM) and the time they take (several days, if not weeks).
The 2D models, although very fast (quarter of an hour to an hour, with few GBs of RAM), were too inaccurate because of
their inability to reflect the axisymmetry that is inherent in the GEM detector. As a result,
they are unable to provide the correct field map even for the central hole of the device. The
2D axisymmetric models have been found to be the most optimum because of their geometry,
ability to provide the correct fields for the central hole, while being less resource hungry
(few GBs of RAM) and reasonably fast (tens of minutes to an hour). In addition, a reasonable scaling factor could be easily derived for these models in order to compensate for the lack of contribution of the non-central holes. From these studies, we have concluded that an axisymmetric model serves the purpose of simulating a GEM-based detector reasonably well.
\section{Results and Discussions}
\label{sec:ResDis}
In this section, the evolution of the electron and ion fluids in a 2D axisymmetric gas volume has been presented.
The initial amounts of electronic and ionic fluids have been determined by computing the energy deposition \cite{Heed} in the drift region due to the passage of the 5.9 keV photons from Fe$^{55}$ radioactive source for the gas mixture Ar + CO$_{2}$ in proportions 70:30 by volume.
The simulations have been carried out for single, double and triple GEM configurations with different geometrical features and applied voltages, as mentioned in the table \ref{tab:table2}.
Choice of this radioactive source and geometry configuration follows the description of \cite{Bachmann:2000az}.

The volume integral of the charge cluster gives the total number of electrons and ions present at each time step of the simulation.
These numbers provide estimates of the total charge multiplication within the detector volume. However, the effective gain is less than this number due to losses at various stages of the detector. The effective gain has been estimated by integrating the signal induced at the anode, and also by computing the total number of electrons flowing through the anode.

The following stopping conditions have, in general, been implemented in the model to identify the avalanche and streamer and to stop the simulation.
\begin{enumerate}
	\item When the total number of electrons in the gas volume have become less than 1, the simulation has been stopped and the mode of operation has been identified as avalanche.
	\item When the total number of electrons in the modelling volume reaches the critical Raether limit of $5 \times10^6$ \cite{Gasik:2017uia}, the simulation has been stopped, identifying the mode of operation as streamer.
\end{enumerate}
However, several studies have been carried out without the second condition, in order to investigate the effectiveness of the upper limit.
\subsection{Primary ionization}
\label{subsec:Primary}
\begin{figure}[htbp]
\centering
\includegraphics[width=0.49\linewidth]{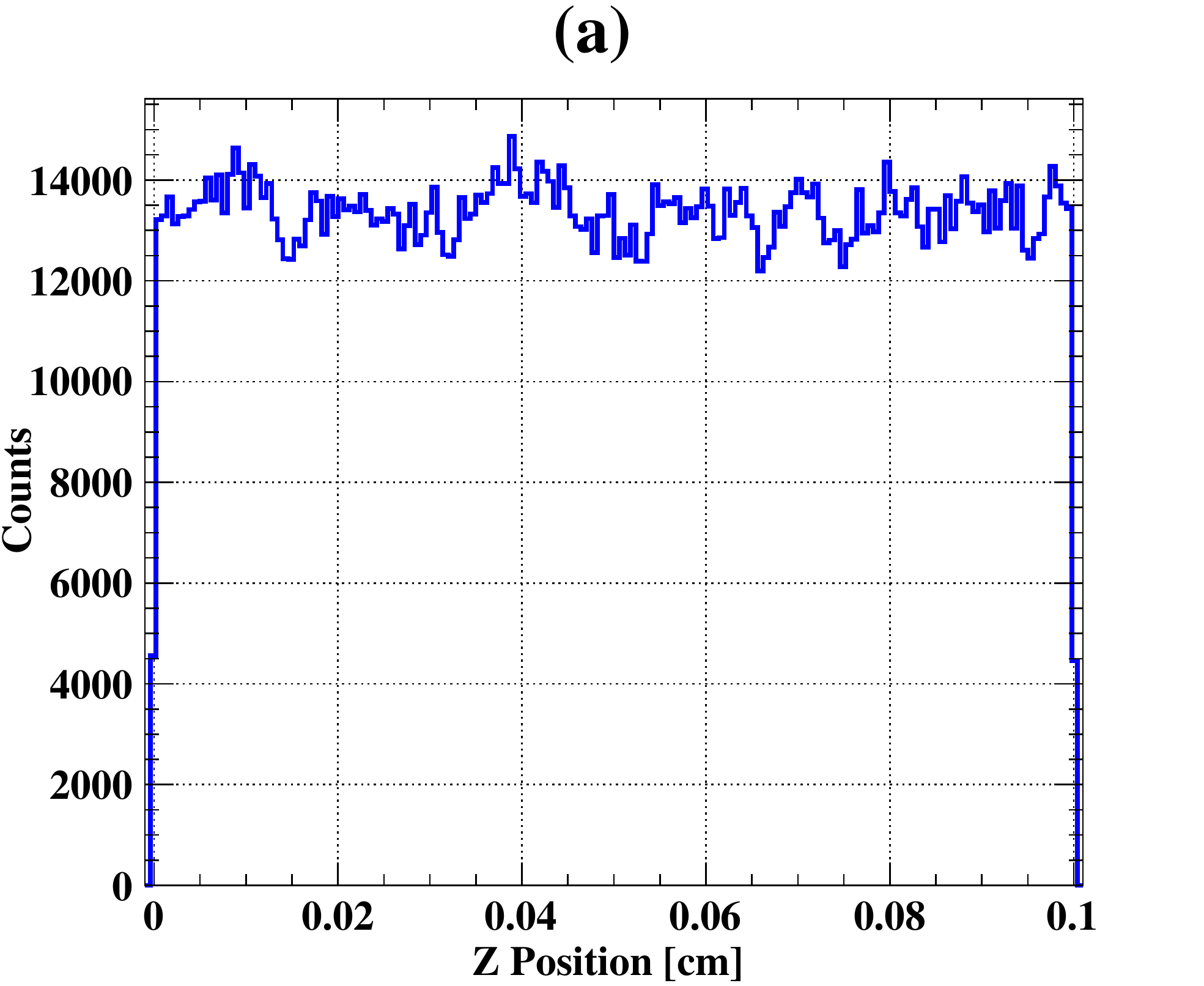}
\includegraphics[width=0.49\linewidth]{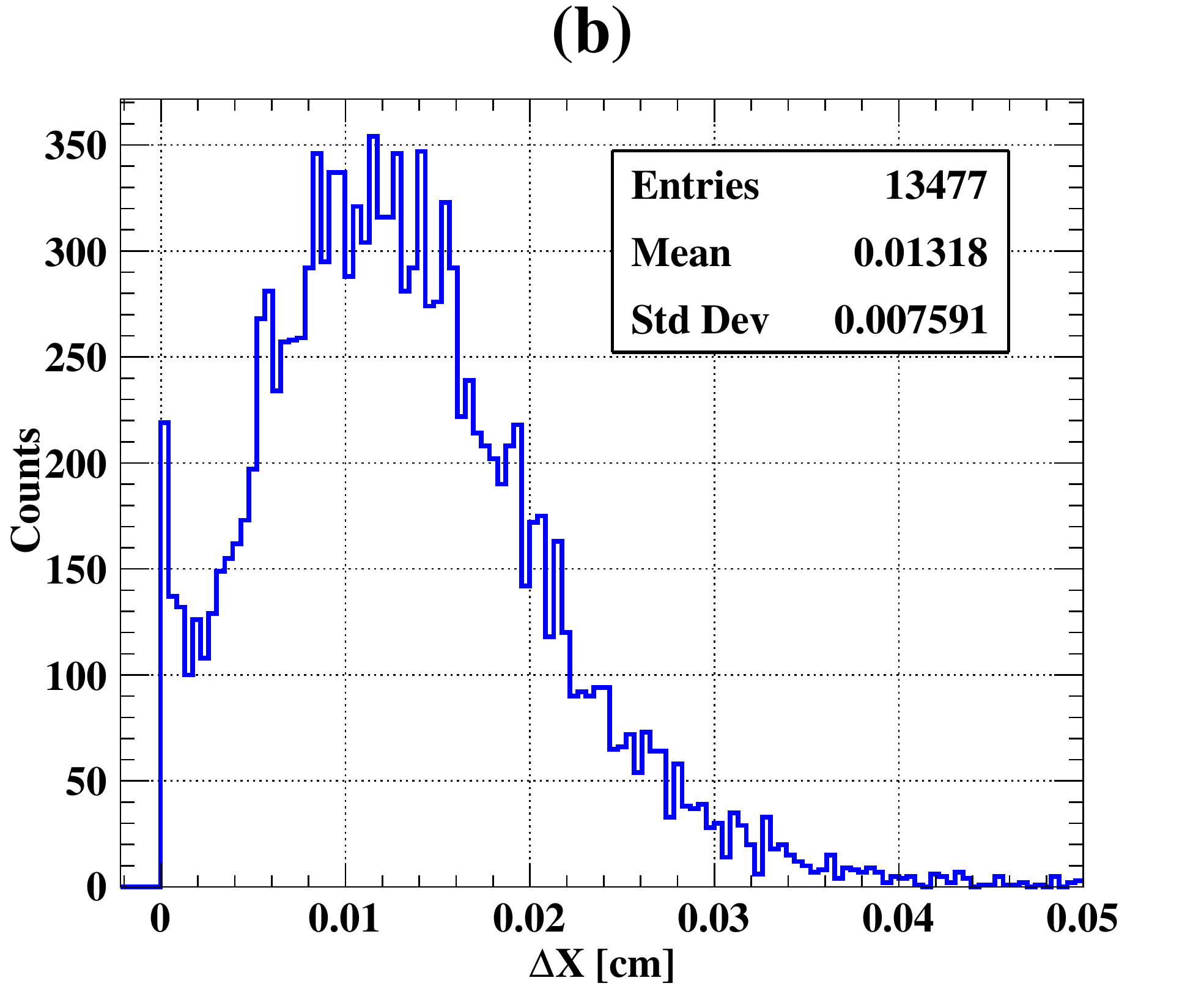} \vskip 4pt
\includegraphics[width=0.49\linewidth]{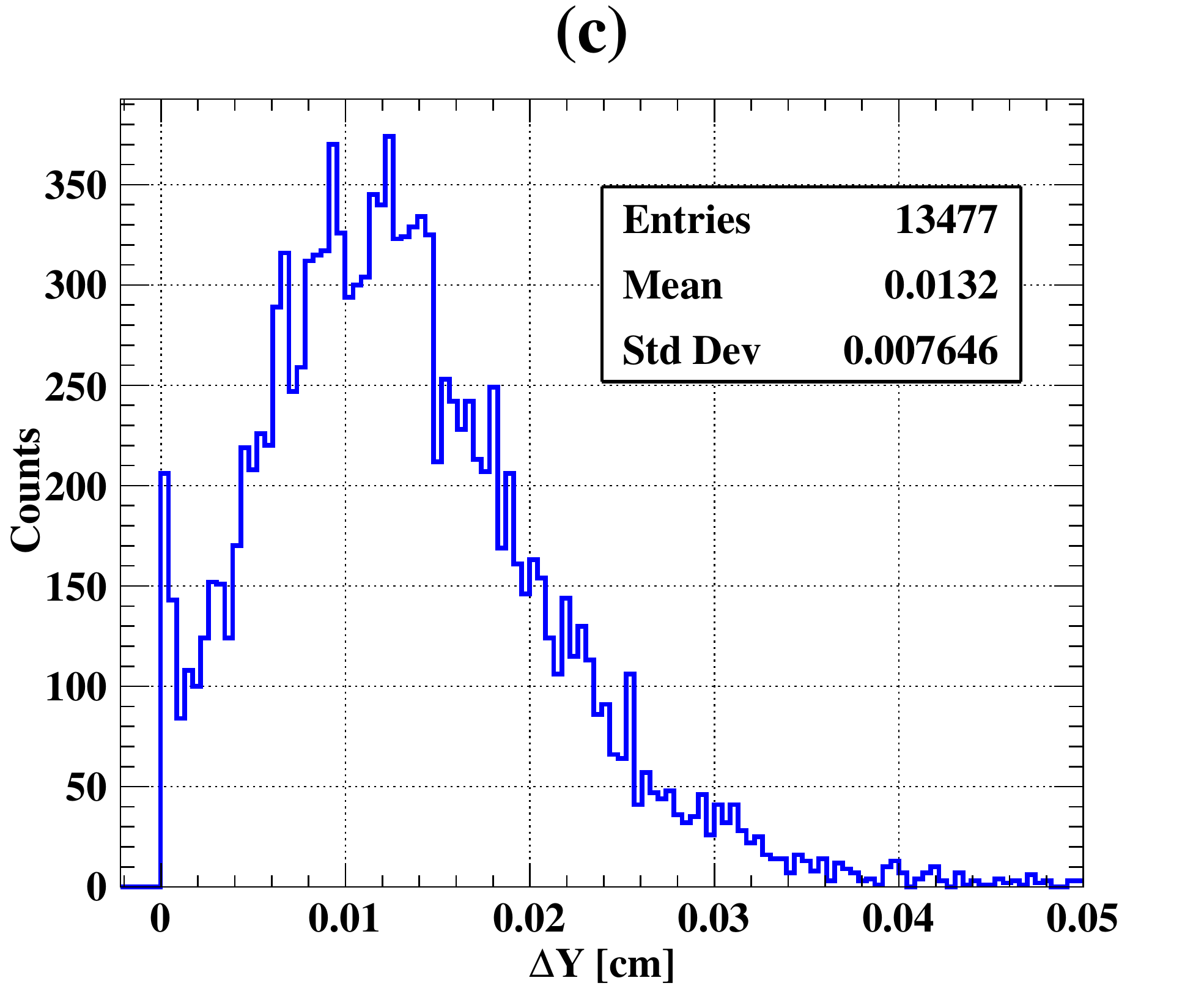}
\includegraphics[width=0.49\linewidth]{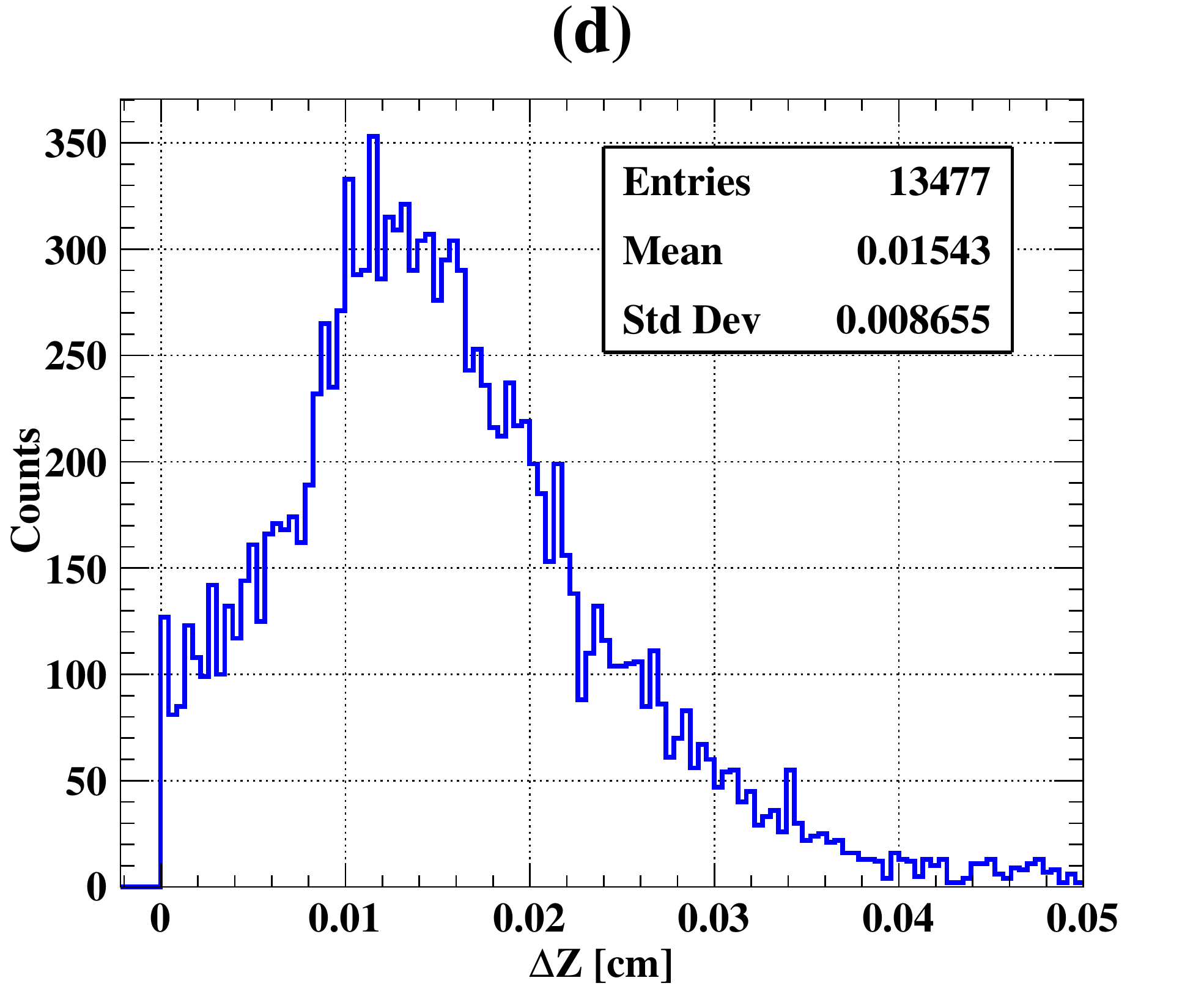}
\caption{ (a) Z-position of primary electrons. (b), (c) and (d), spread of primary cluster along X, Y and Z-directions respectively.}
\label{fig:fig4}
\end{figure}

In figure \ref{fig:fig4}(a), we have plotted the Z-position of the primary electrons of each event for the Fe$^{55}$ radiation source. The Z-direction span of the gas gap is 1mm. The incident photon is released into the gas volume from the top of the drift volume. The cluster width of each event for the Fe$^{55}$ source along X, Y and Z direction has been shown in the figures \ref{fig:fig4}(b), (c) and (d) respectively. The spread for each event, in each direction have been calculated by subtracting the maximum of each coordinate from its minimum value.
From the figures \ref{fig:fig4}(b), (c) and (d), it is evident that for Z-direction, the axial direction in case of 2D axisymmetric model, the cluster has mean spread of 154 $\mu m$. For both X and Y-directions, radial direction in case of 2D axisymmetric model, cluster has mean spread of 132 $\mu m$. The standard deviation of the seed cluster has been chosen such a way that the 5$\sigma$ \cite{Jaydeep} of the Gaussian primary electron distribution does not go beyond the aforementioned mean spread in any of the directions.

The electric field lines in the drift region collimate the electrons towards the central hole, but due to diffusion a significant amount of primary electrons can be lost to the copper foil or move to the non-central holes \cite{Roy:2020gwl}. Keeping this in mind we have chosen the mean Z-position of the seed cluster to be 250 $\mu m$ above the top GEM foil so that avalanche and streamer mode operations can be easily simulated. The seed position affects the subsequent evolution in various ways, for example through change
in drift time, modification of electron diffusion leading to different charge sharing patterns. In the present work, we have been specially interested in exploring the possibility of simulating the transition from avalanche to streamer modes in GEM-based detectors. This transition is more probable when the seed is reasonably close to the GEM foil. This is reason why the Z-position of
the seed has been maintained at 250 $\mu m$  for all the simulations presented in this work.
\subsection{Avalanche mode operation}
\label{subsec:Ava}
\begin{figure}[htbp]
\centering
\includegraphics[width=0.7\linewidth]{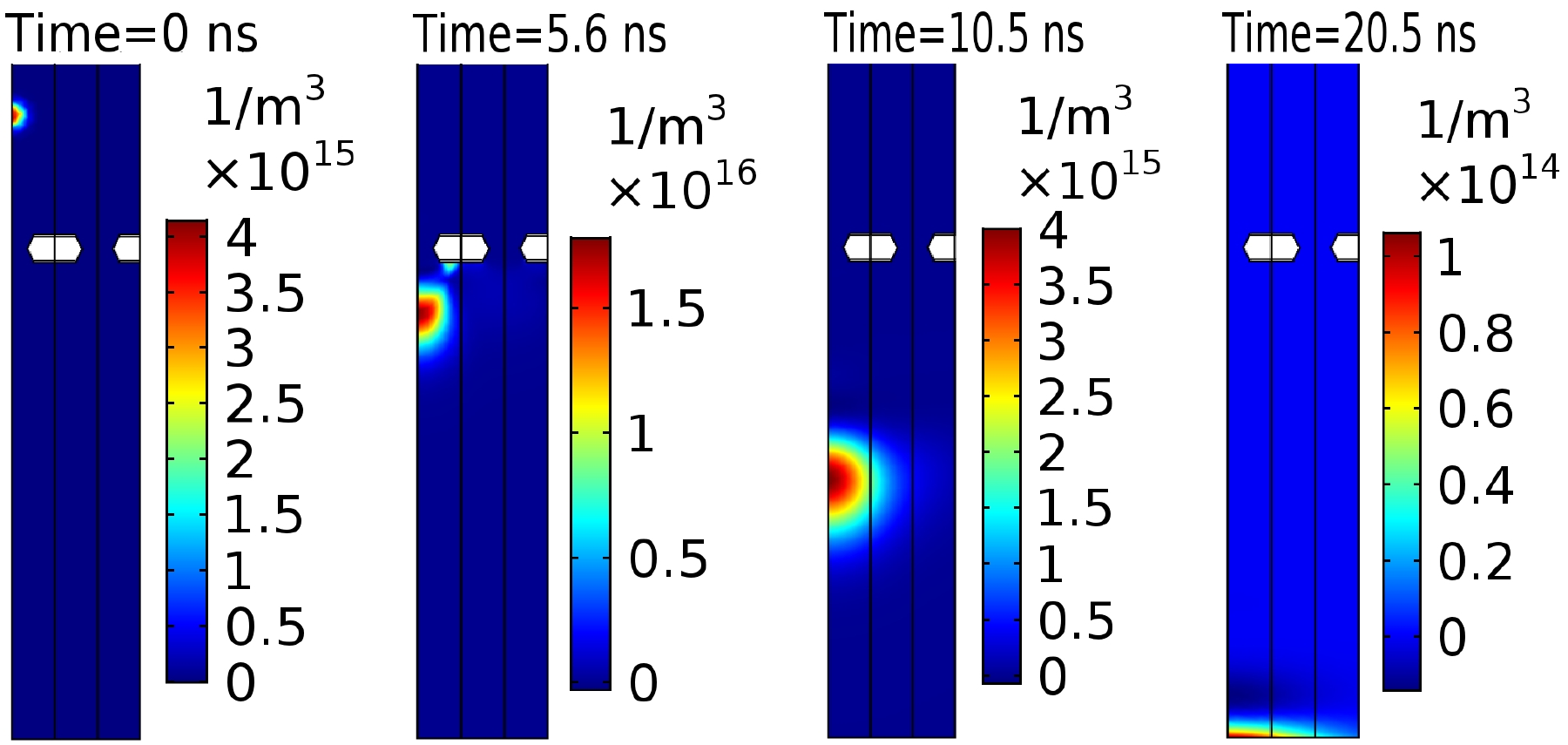}
\caption{Avalanche formation in a single GEM at $\Delta V_{\rm GEM} = 500V$}
\label{fig:fig5}
\end{figure}

\begin{figure}[htbp]
\centering
\includegraphics[width=0.49\linewidth]{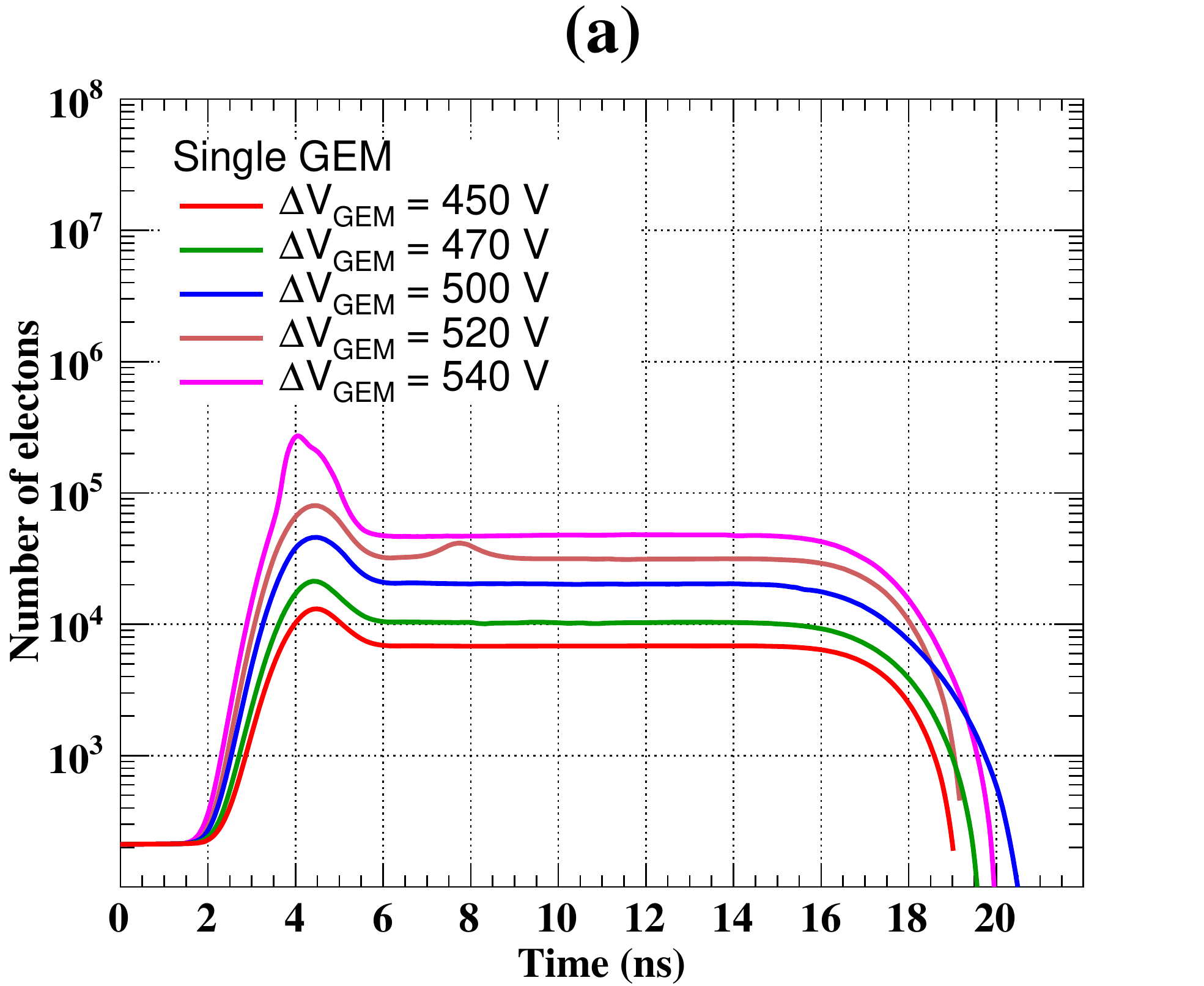}
\includegraphics[width=0.49\linewidth]{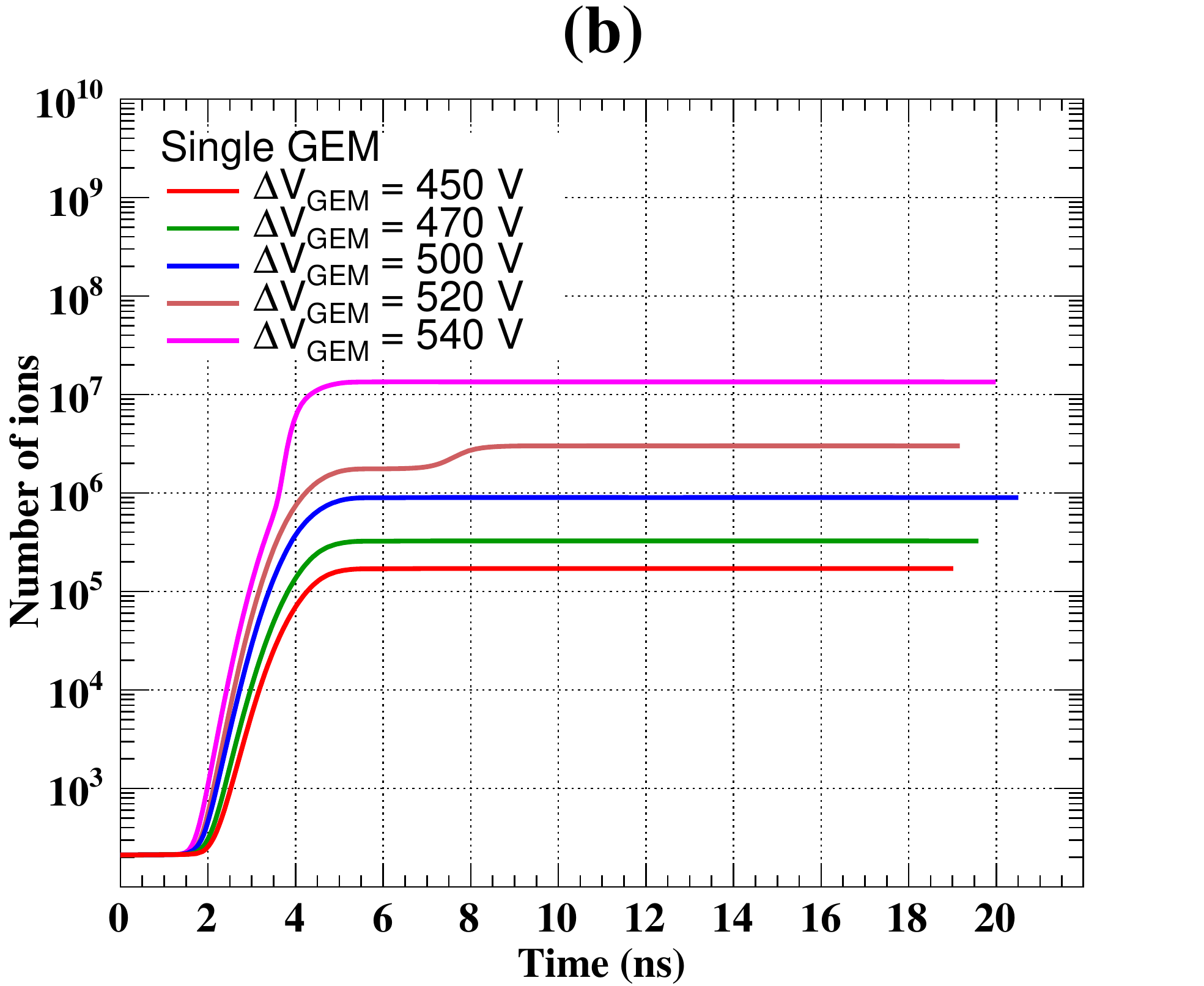}\vskip 4pt
\includegraphics[width=0.49\linewidth]{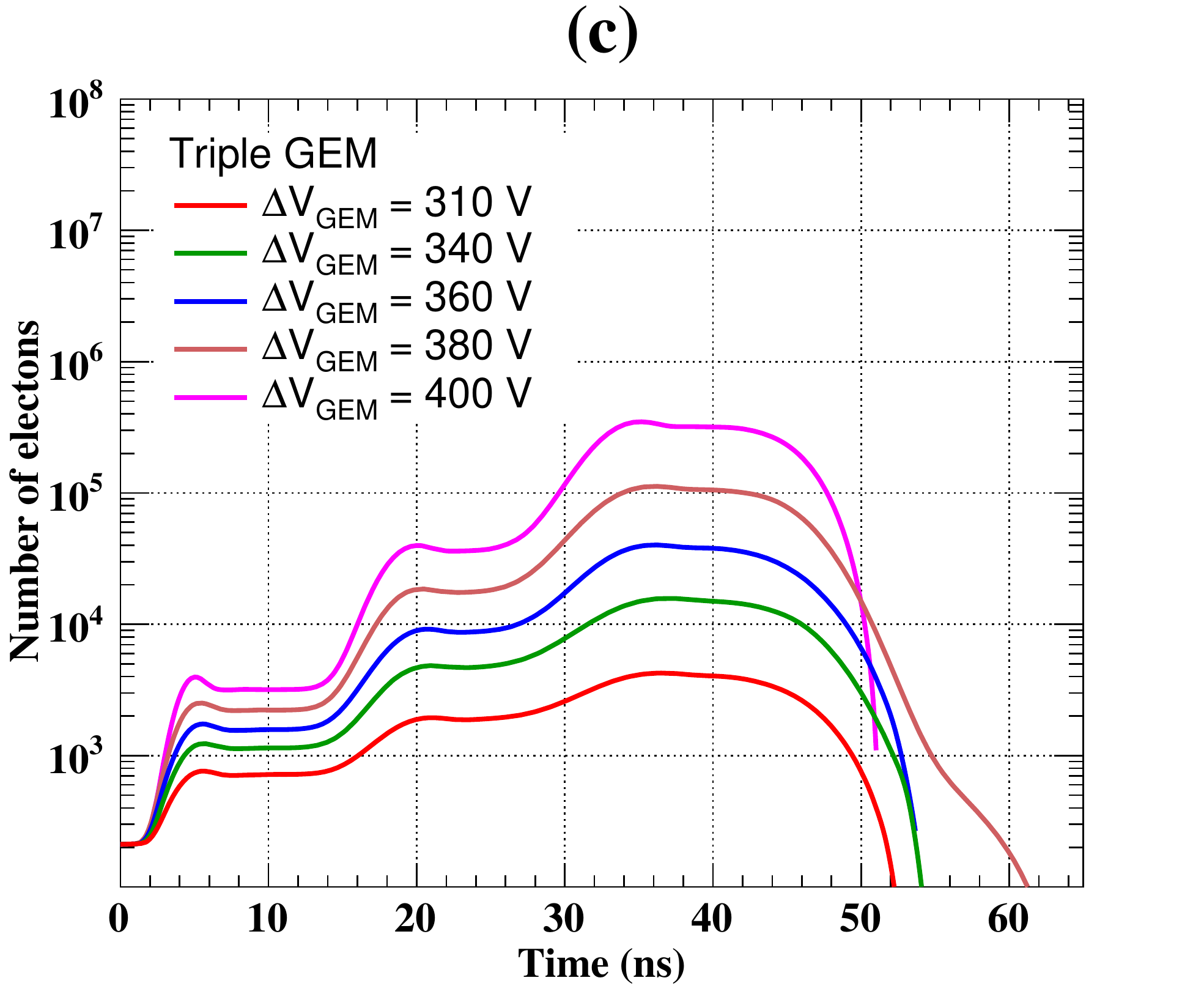}
\includegraphics[width=0.49\linewidth]{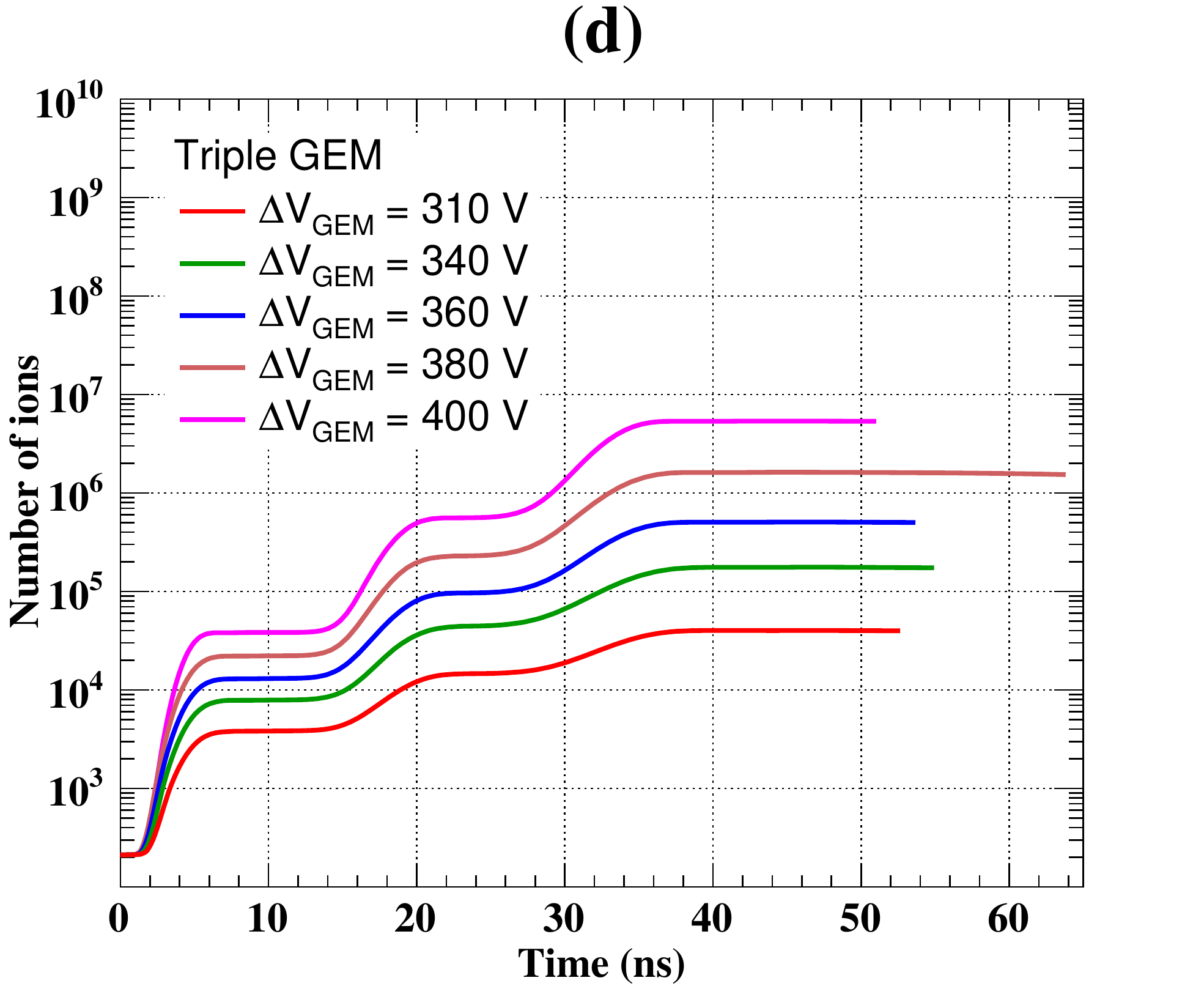}
\caption{(a) and (b) evolution of electrons and ions for a single GEM. (c) and (d) evolution of electrons and ions for a triple GEM}
\label{fig:fig6}
\end{figure}

In the avalanche mode operation of the detector, the electron avalanche proceeds in the gas volume until it reaches the anode, as shown in figure \ref{fig:fig5}.
The evolution of total number of electrons and ions with different voltages applied across the GEM foil for a single GEM  has been shown in figures \ref{fig:fig6}(a) and (b) respectively.

Similarly, figures \ref{fig:fig6}(c) and (d), show the evolution of electrons and ions in the detector volume for a triple GEM detector with different voltages applied across the GEM foils.
The three amplification stages are clearly visible in these two figures. The total number of electrons and ions are not very different for single and triple GEM detectors, but the amplification per stage of a triple GEM detector is significantly less than that of a single GEM detector for obvious reasons.

\begin{figure}[htbp]
\centering
\includegraphics[width=0.49\linewidth]{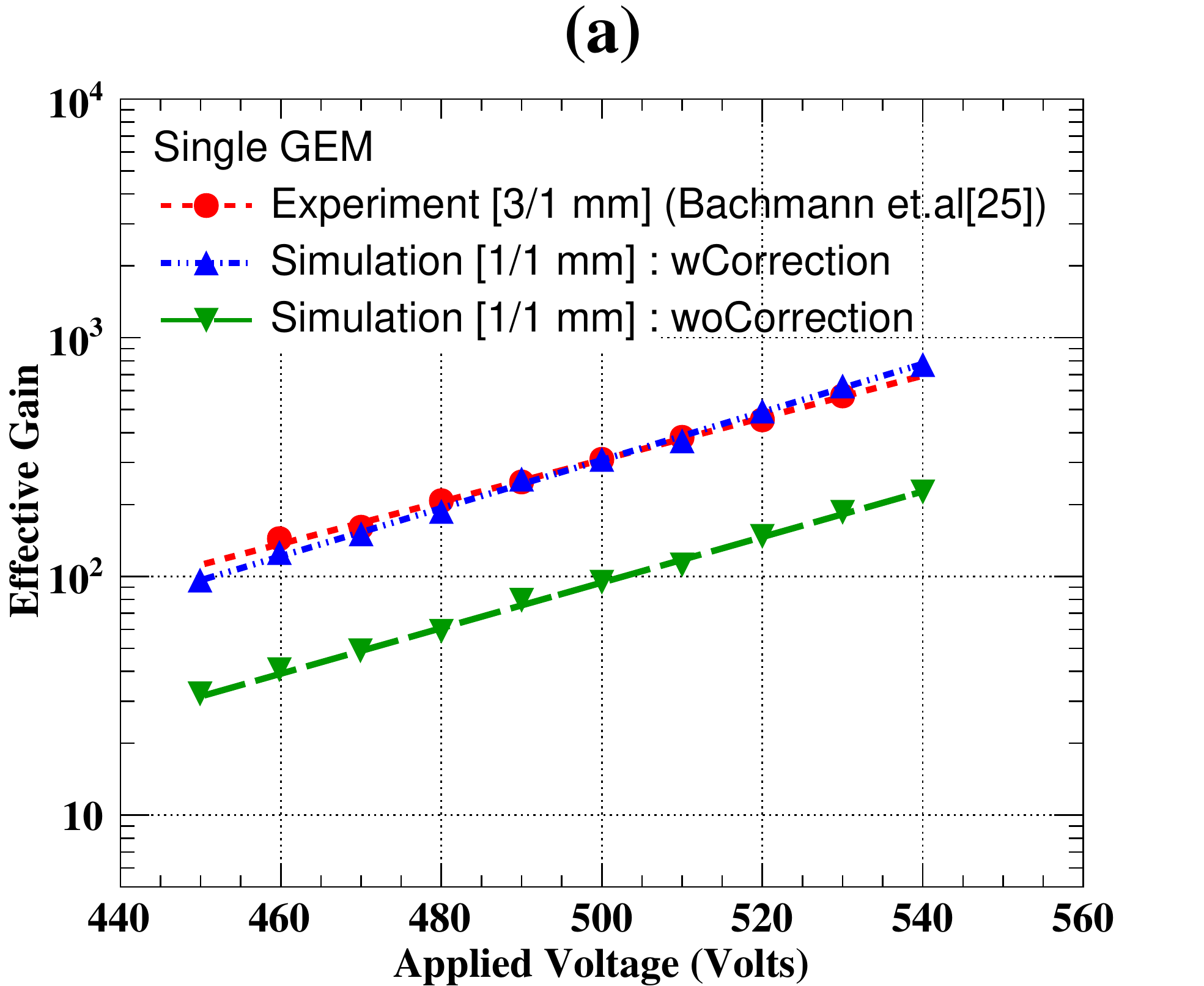}
\includegraphics[width=0.49\linewidth]{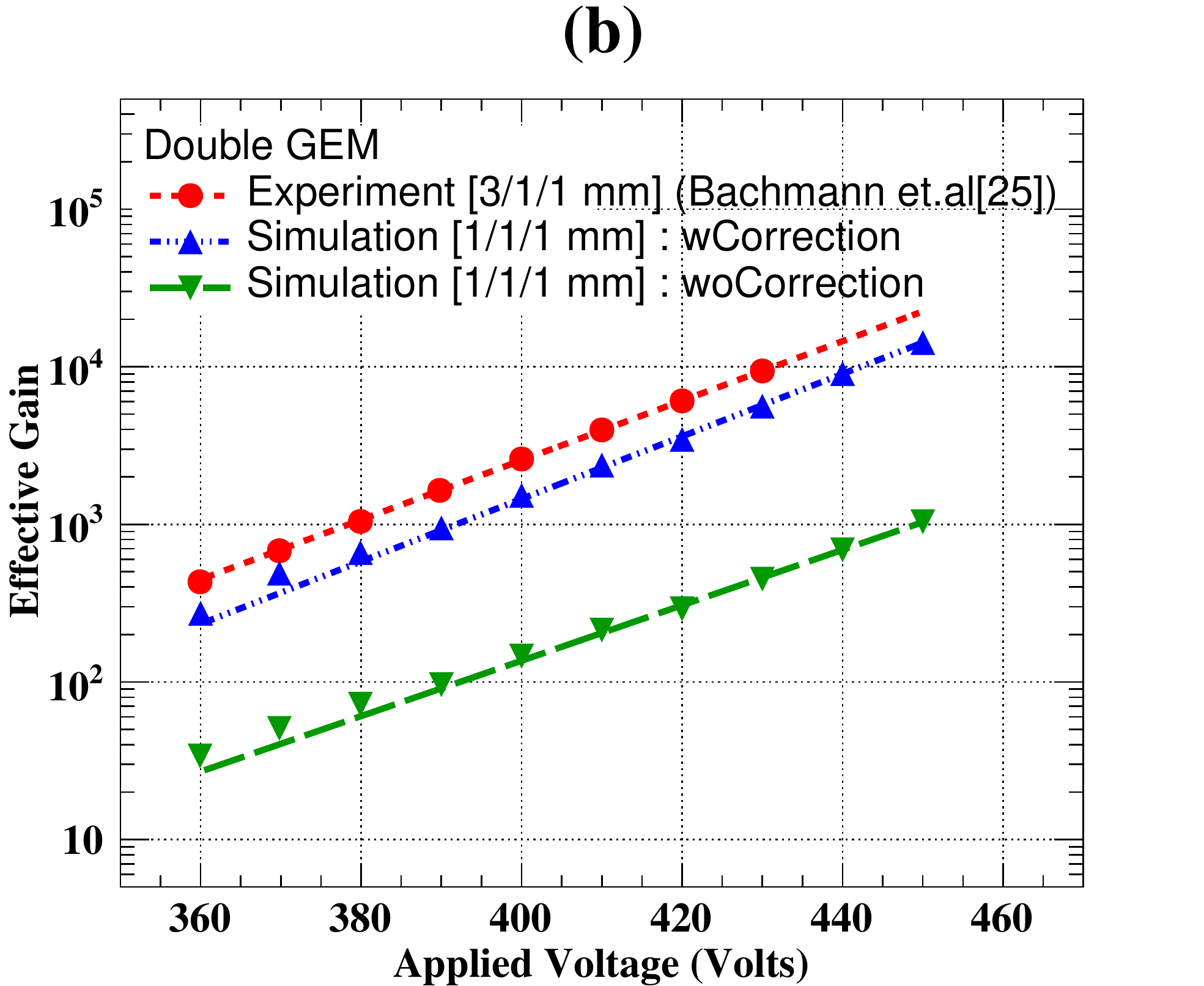}
\caption{Comparison between experimental gain reproduced from \cite{Bachmann:2000az} and numerical estimate for single GEM in (a) and double GEM detector in (b). The dotted line represents the fit of the experimental and simulated gain.}
\label{fig:fig7}
\end{figure}

\begin{figure}[htbp]
\includegraphics[width=0.49\linewidth]{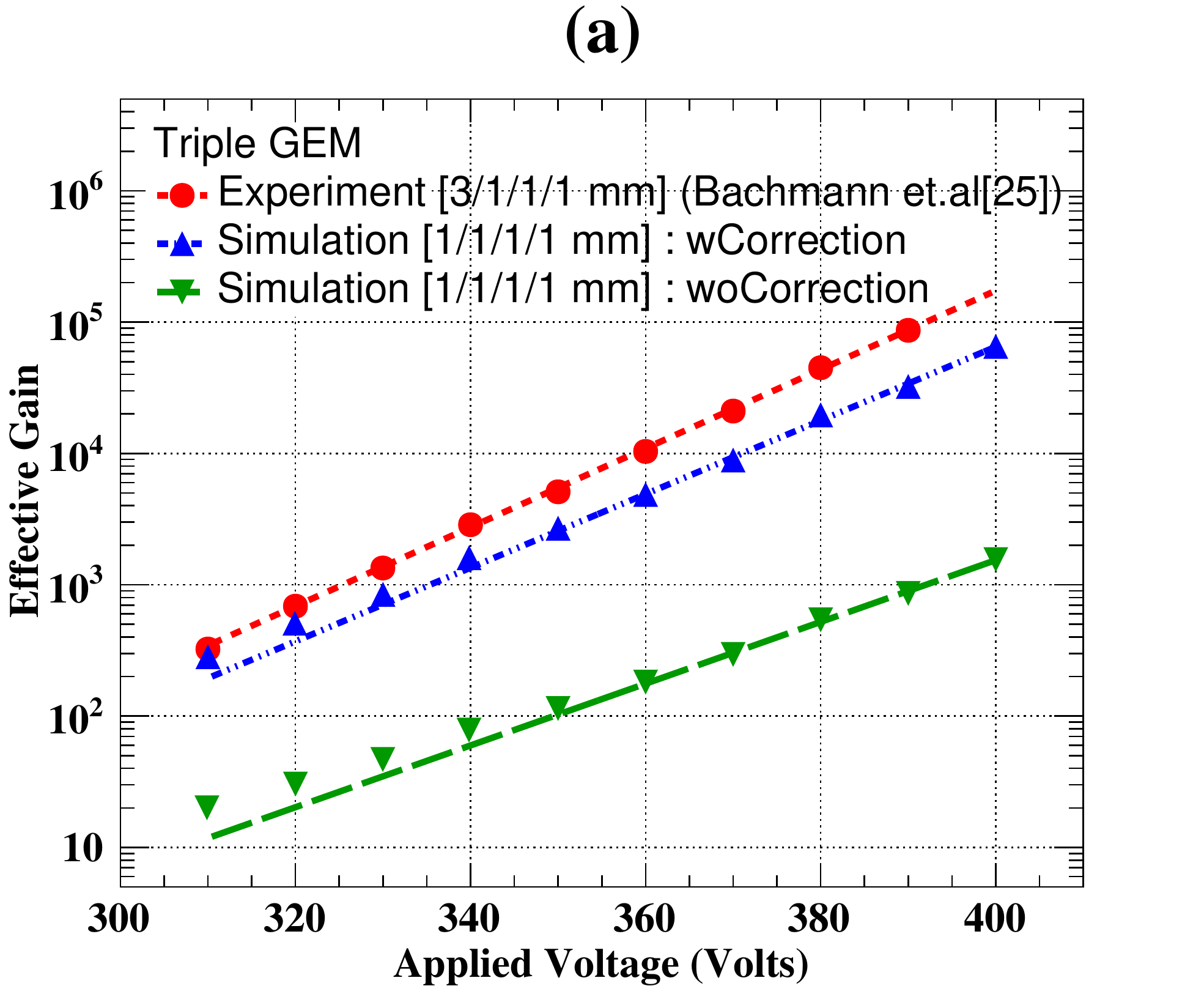}
\includegraphics[width=0.49\linewidth]{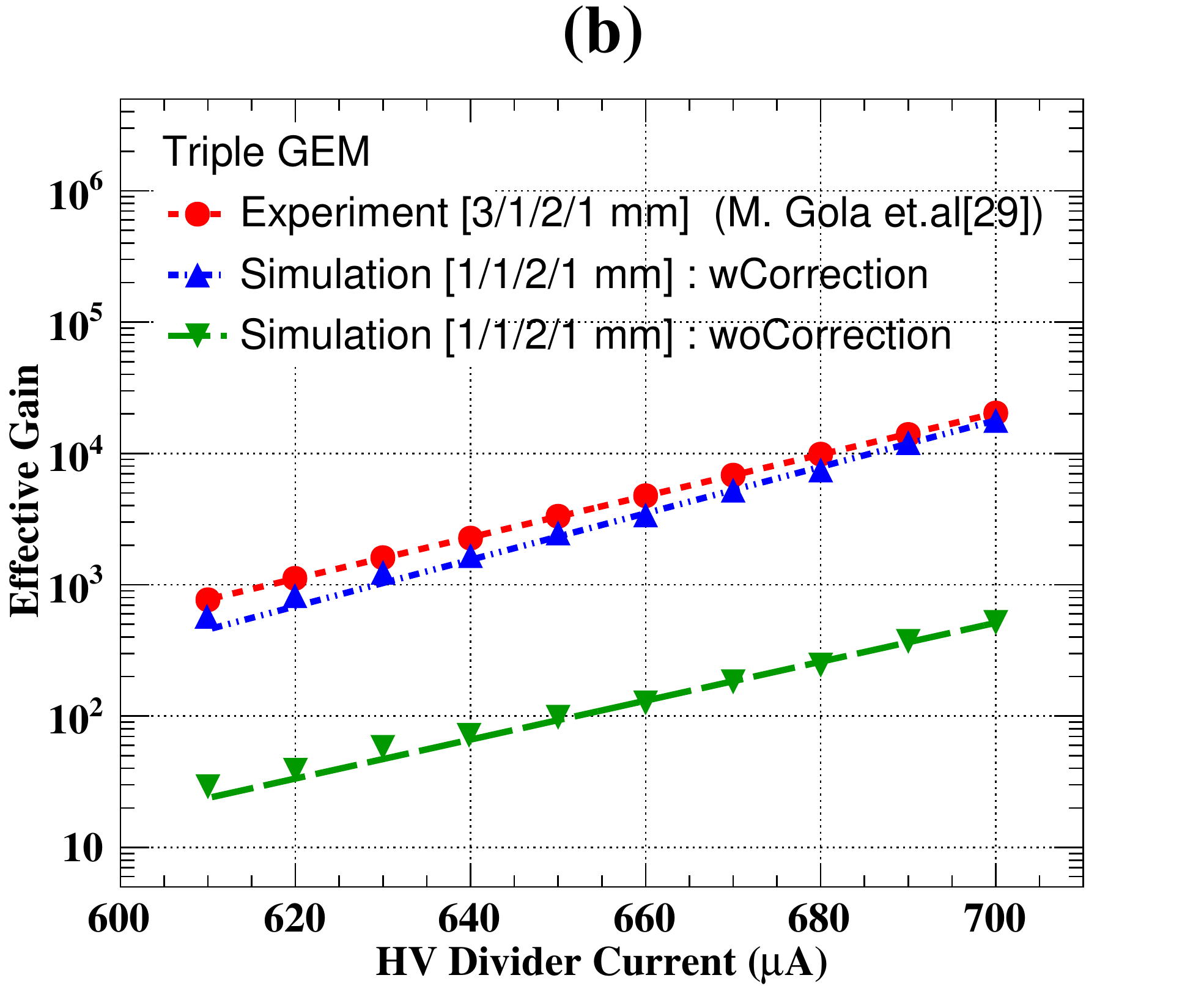}
\caption{(a) Comparison between experimental gain reproduced from \cite{Bachmann:2000az} and numerical estimate for the triple GEM detector in 1/1/1/1 gas gap configuration. (b) Comparison of experimental gain reproduced from \cite{Gola:2018njy} and the numerical estimate for the triple GEM detector in 1/1/2/1 gas gap configuration. The dotted line represents the fit of the experimental and simulated gain.}
\label{fig:fig8}
\end{figure}

As mentioned earlier, while the plots indicate the total amount of multiplication within the detector, it does not quantify the effective gains obtained from these detectors at these parameter points.
For estimating the effective gain, the number of electrons collected at the anode has been estimated by using a boundary probe that measures the flux of electrons through the electrode.
The integration over time of this flux is also a measure of the effective gain after an appropriate division by the number of primaries at the initial moment.
Using this method, the gain of the detectors considered in this work have been estimated and compared with experimentally observed gain from \cite{Bachmann:2000az} in figures \ref{fig:fig7}(a), (b) and \ref{fig:fig8}(a).
It is observed that while the trend of the effective gains are comparable for all three detector configurations, experimentally and uncorrected numerical estimates (woCorrection), the values are significantly different. In the section \ref{subsec:Electric Field} we have shown in figures \ref{fig:fig2}(a) and (b), that the electric field of the central hole matches with that observed in 3D, but the same in the non-central holes differs by 20\%. This reduction in electric field reduces the gain for non-central holes.
Moreover, \cite{Roy:2020gwl} and other similar works report that significant number of electrons goes to the other holes.
These two factors together can explain the mismatch of gain between simulation and experiment to some extent.
A correction factor has been calculated using the field value and charge sharing information. Inclusion of this correction factor improves the simulated gain (wCorrection) as shown in figures \ref{fig:fig7}(a), (b) and \ref{fig:fig8}(a). As can be seen from Appendix \ref{sec:Appendix 1}, computation of the correction factors have been carried out in a rather simplistic manner, true to the nature of a fast simulator. A more elaborate and diligent approach towards computing these factors is expected to further improve the agreement.

The simulation is also performed with a double-mask triple GEM detector consisting of 1 mm drift gap, 1 mm transfer - 1, 2 mm transfer - 2 and 1 mm induction gap. It may be noted here that, similar gas gaps are also utilized for the triple GEM detector being used to upgrade in the muon system of the CMS experiment \cite{CMS2015}. The numerical estimate of the gain with and without correction applied from this simulation is compared with the experimentally observed gain from \cite{Gola:2018njy} in figure \ref{fig:fig8}(b).   

It may be noted here that, the drift gap considered in the experiments \cite{Bachmann:2000az} and  \cite{Gola:2018njy} is 3 mm. However, the same electric field is maintained across the drift gap for both in simulation and experiment. Therefore, the choice of 1 mm drift gap in the simulation is reasonable instead of 3 mm as in the aforementioned experiments.
\subsection{Effect of space charge and streamer formation}
\label{subsec:ST}
Streamer formation occurs when the field produced by the space charges becomes comparable to the externally applied field in the gas volume \cite{Raizer}.
Its development depends on several factors such as the applied voltage, gas pressure, the field gradients, the geometry and curvatures of the electrodes, presence of humidity and possible dielectric surfaces.
These factors may lead to the formation of either positive and negative streamers in the gas volume.
The electron avalanche is known to undergo transition into a negative streamer propagating towards anode when the space charge density in the avalanche front reaches the critical number needed for the streamer to initiate.
The positive streamer, on the other hand, occurs when the density of ions left behind in the gas volume grows sufficiently high.
The resulting streamer propagates towards the cathode generating even higher densities of ions and electrons.

\begin{figure}[!htbp]
\centering
\includegraphics[width=0.49\linewidth]{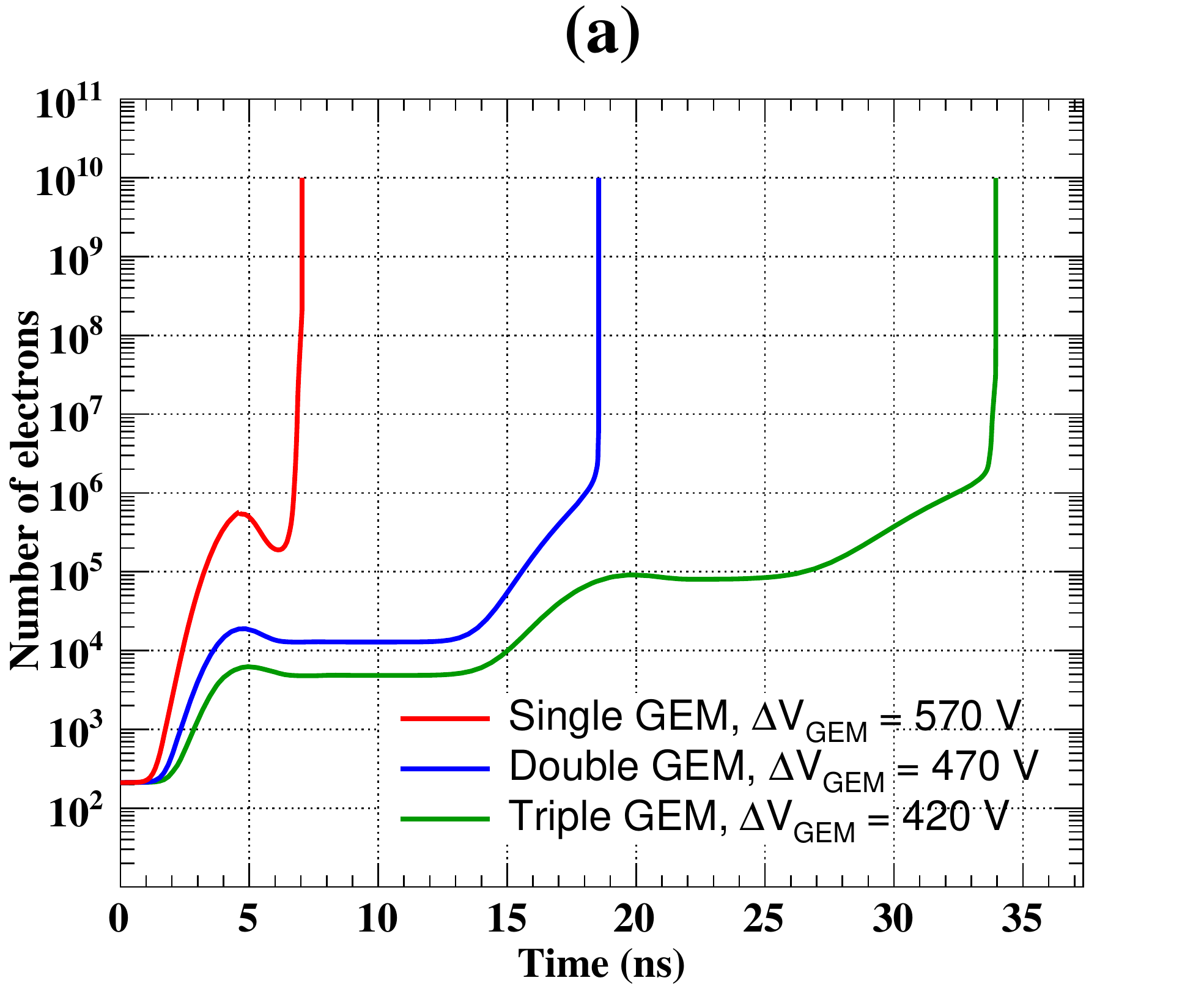}
\includegraphics[width=0.49\linewidth]{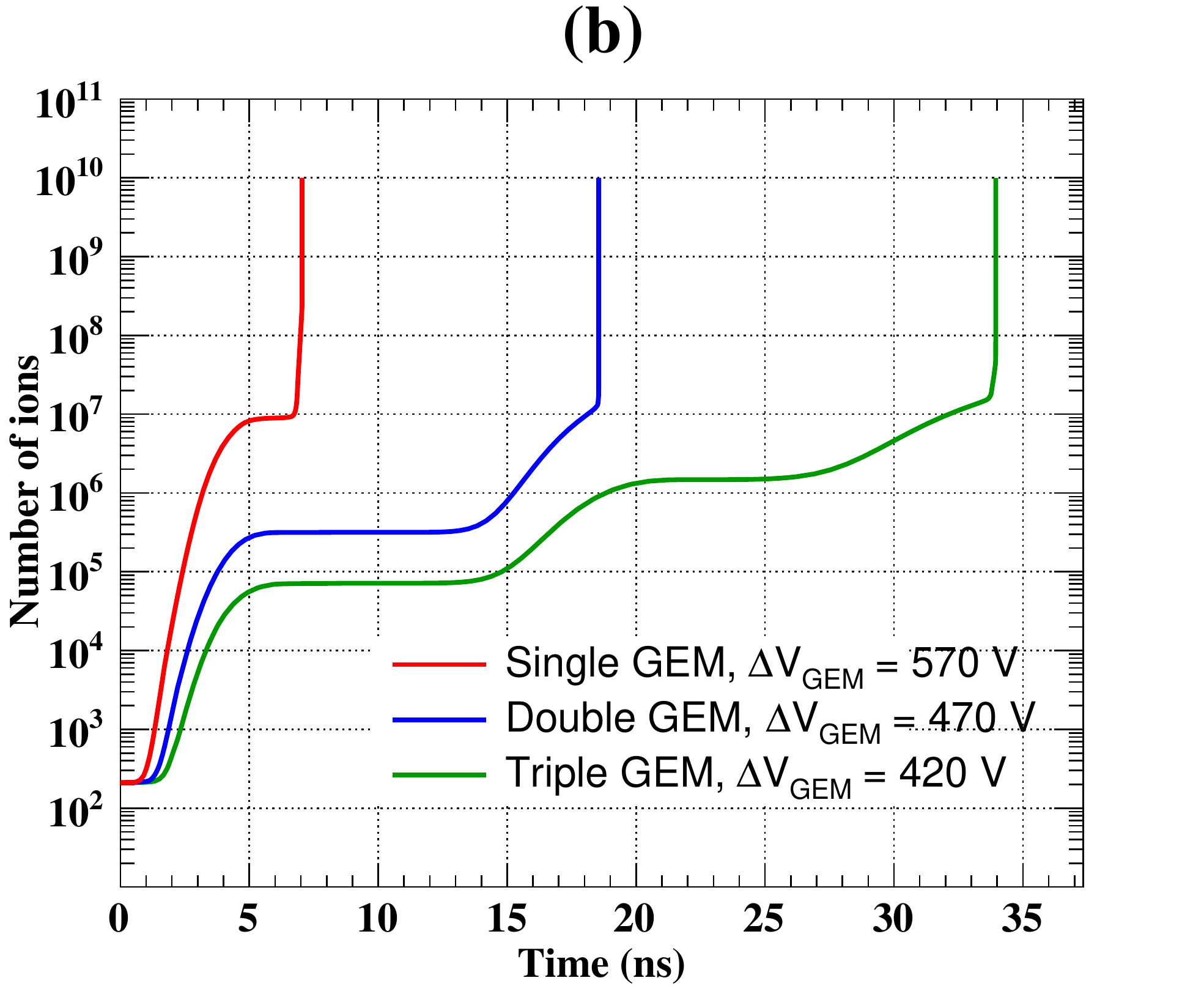}
\caption{Evolution of electron numbers in (a) and ion numbers in (b) during transition to streamer for single, double and triple GEM detectors.}
\label{fig:fig9}
\end{figure}

In the present work, the positive streamers have been observed to initiate close to the GEM anode.
The transition to the streamer mode, as shown in figures \ref{fig:fig9}(a) and (b), occurs when the total number of electrons and ions in the detector volume approaches $10^6$ and $10^7$ in the detector volume, respectively.
For a single GEM, the transition occurs at a lower value of total number of electrons ($8 \times 10^5$), while for double and triple GEM detectors, it occurs close to $2 \times 10^6$.
It may be mentioned here that similar values for streamer transition in GEM-based detectors have been reported experimentally by \cite{Gasik:2017uia}.

The transition to streamer for a single GEM can be compared with experimental trace \cite{Raether1964} and hydrodynamic simulation for parallel plate device \cite{Bhattacharya2020}.
The plots from all the sources share a strong similarity.

\begin{figure}[htbp]
\centering
\includegraphics[width=0.9\linewidth]{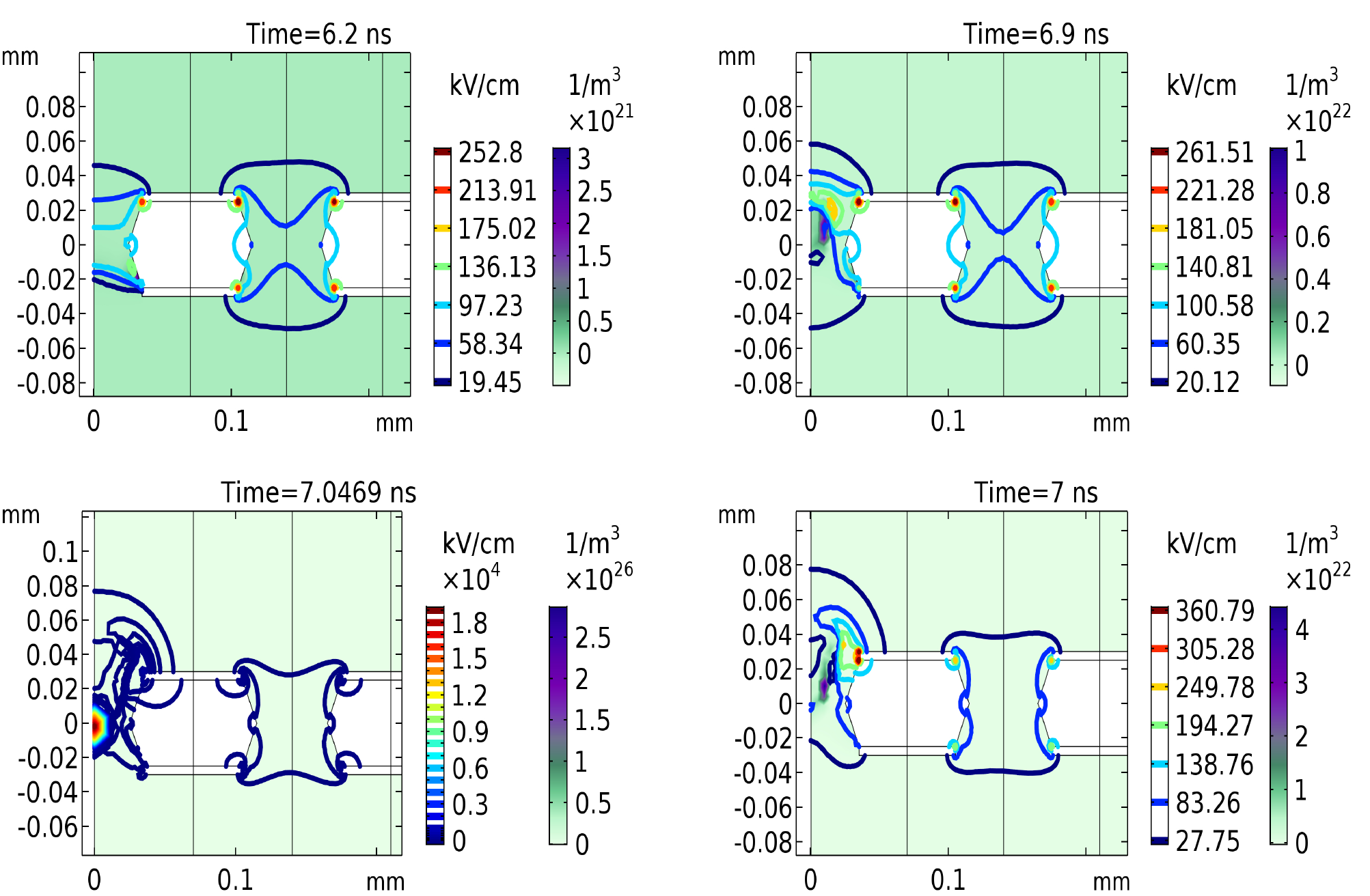}
\caption{Distortion of total electric field due to space charge and evolution of ion concentration during streamer formation in a single GEM at $\Delta V_{\rm GEM} = 570V$ in clockwise direction.}
\label{fig:fig10}
\end{figure}

\begin{figure}[htbp]
\centering
\includegraphics[width=0.9\linewidth]{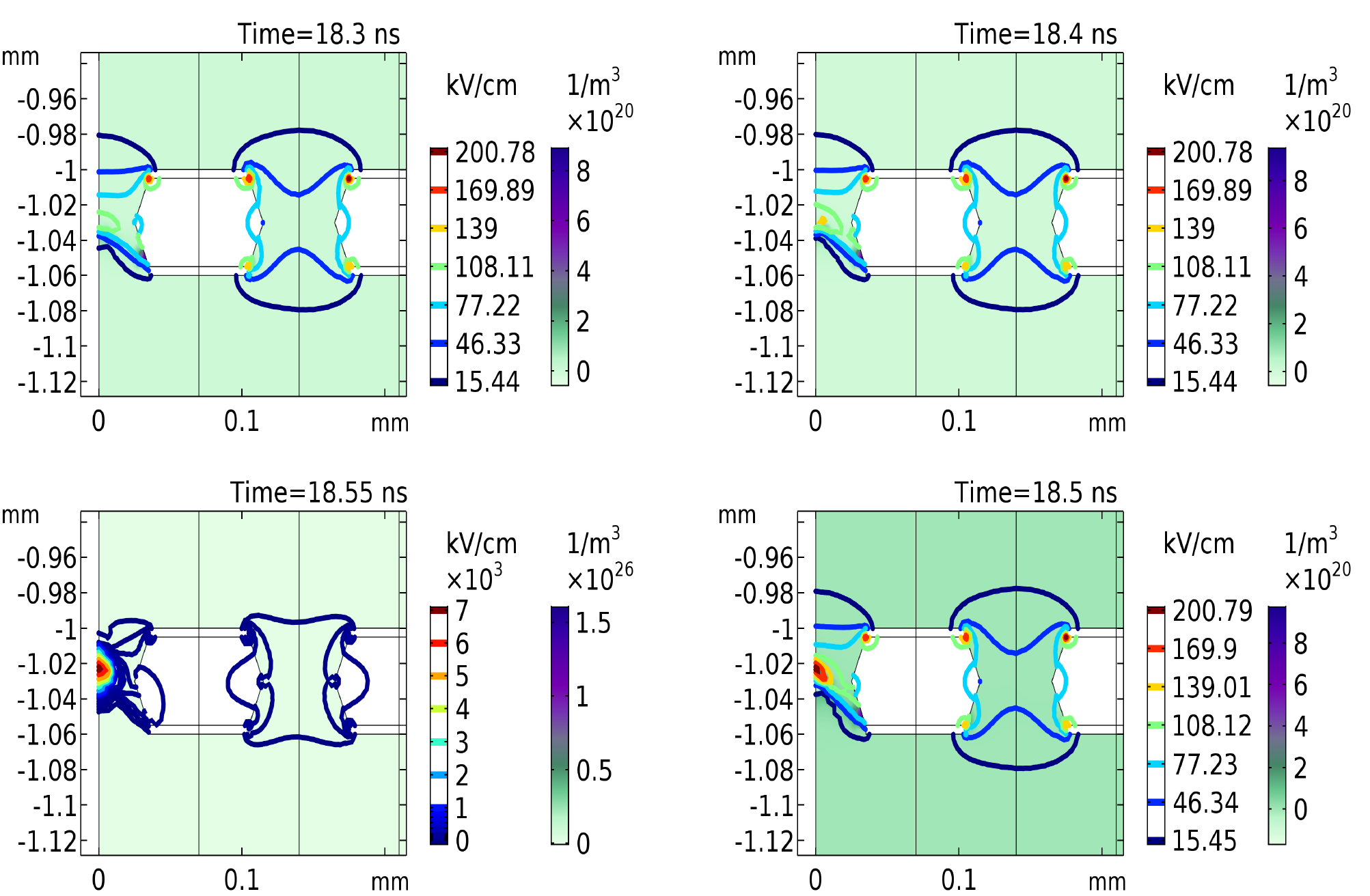}
\caption{Distortion of total electric field due to space charge and evolution of ion concentration during streamer formation in a double GEM at $\Delta V_{\rm GEM} = 470V$ in clockwise direction.}
\label{fig:fig11}
\end{figure}

\begin{figure}[htbp]
\centering
\includegraphics[width=0.9\linewidth]{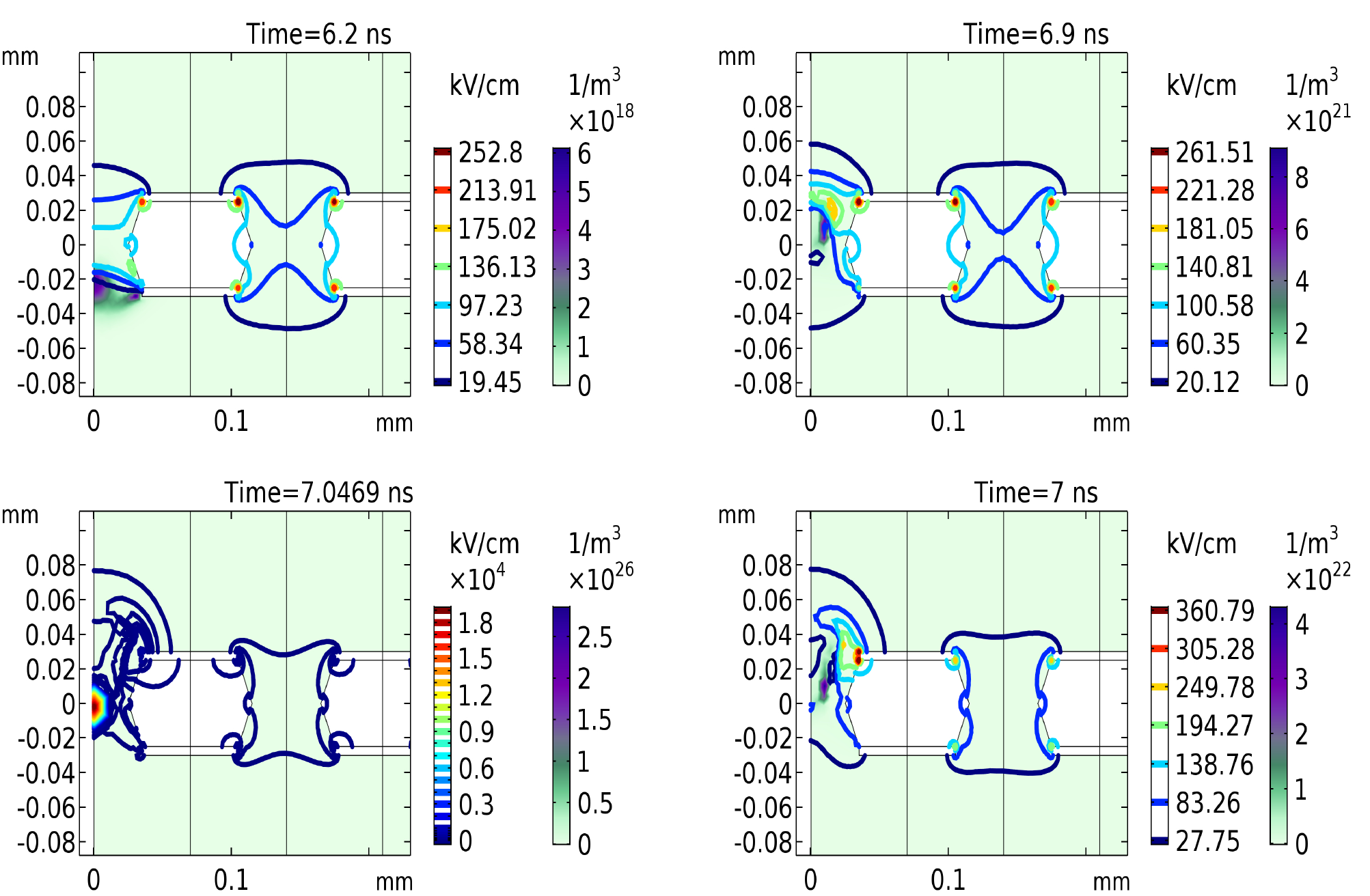}
\caption{Distortion of total electric field due to space charge and evolution of electron concentration during streamer formation in a single GEM at $\Delta V_{\rm GEM} = 570V$ in clockwise direction.}
\label{fig:fig12}
\end{figure}

\begin{figure}[htbp]
\centering
\includegraphics[width=0.9\linewidth]{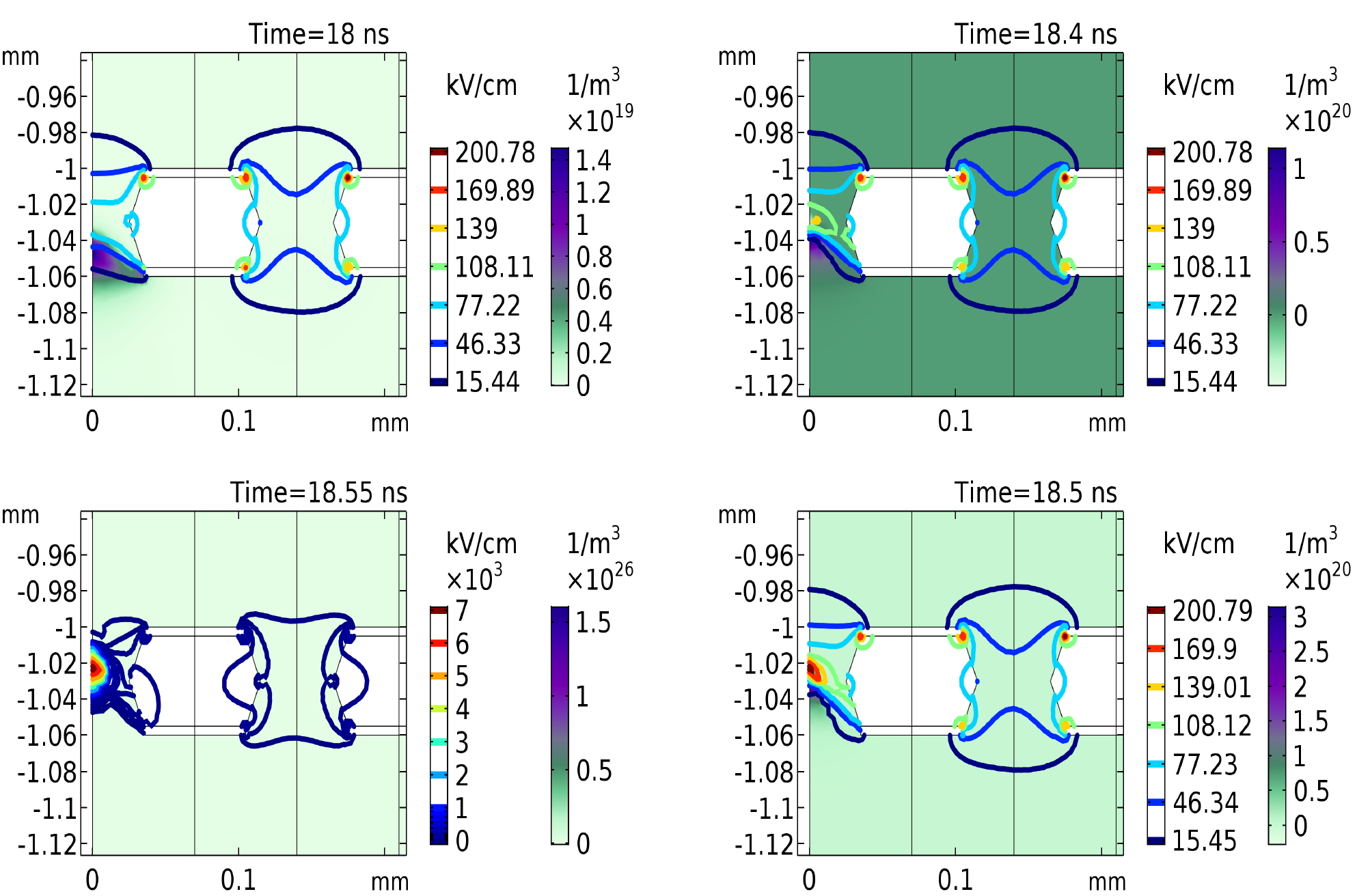}
\caption{Distortion of total electric field due to space charge and evolution of electron concentration during streamer formation in a double GEM at $\Delta V_{\rm GEM} = 470V$ in clockwise direction.}
\label{fig:fig13}
\end{figure}

As is seen from figures \ref{fig:fig10} and \ref{fig:fig11}, the ion density builds up near the GEM anode before moving towards GEM cathode.
The density builds up and the movement is associated with strong distortion of the electric field due to space charge.
For both single (figure \ref{fig:fig10}) and bottom GEM foil of a double GEM (figure \ref{fig:fig11}), the ions are found to get produced in large numbers near the edges of the GEM holes that face the readout anode.
As a result, the applied electric field is found to be distorted significantly (from the initial value of 150 kV/cm to more than 200 kV/cm during the initial stages and close to of the order of thousands kV/cm during the final stages) due to the accumulation of the ionic space charge in this region. 
This large field leads to further rapid multiplication in this region.
Since the process is dominated by the ions, the product of these multiplications (both ions and electrons) are found to propagate towards the GEM cathode in the form of a positive streamer head. Figures \ref{fig:fig12} and \ref{fig:fig13} gives the movement of electron density from GEM anode to GEM cathode for single and double GEM under this distorted electric field respectively.

Finally, it needs to be mentioned that the transition to streamer is not sharp.
Rather it happens over a range of the operating parameter.
For example, for single GEMs, the transition is found to occur in between 540-570 volts, for double GEMs, the range is 450-470 volts while for triple GEMs it is found to be 400-420 volts.
In these voltage ranges, seemingly transition regions, significant oscillation is observed in the solution.
\section{Conclusion}
\label{sec:Conc}
In this paper, we have proposed a fast hydrodynamic simulation framework to study avalanche and streamer evolution for gaseous
detectors based on GEM. This framework has been used to simulate single, double and triple GEM detectors under various operating
conditions. The computationally difficult transition from avalanche to streamer mode has been modeled with a fair amount of success.

Systematic studies on various aspects of a number of mathematical / numerical possibilities have been carried out to find an optimum
set of parameters that preserves the capability of modeling major physics processes with reasonable accuracy, while being
computationally affordable in a present day laboratory. Requirement of scaling of charge amplification from 2D
axisymmetric to full 3D models has been noted here.

Simulations of single, double and triple GEM geometries as described in \cite{Bachmann:2000az} have been performed. The voltage range
for occurrence of streamer and avalanche matches what described in \cite{Bachmann:2000az}. The simulated gain, which was initially
found to be less, agreed with experimental data reasonably well after it was scaled using factors derived following the method given in Appendix \ref{sec:Appendix 1}. The
effect of space charge in the distortion of the applied electric field and in the growth of avalanche and streamer discharges have
been closely monitored. The evolution of positive streamers dominated by ion transport from GEM anode to GEM cathode have been clearly
observed. Estimates of discharge limits from the simulations presented here are found to be close to the critical discharge limit as
mentioned in \cite{Gasik:2017uia}, \cite{Raether1964}. It may be concluded from the above studies that the proposed model is capable
of describing the charge dynamics in GEM-based detectors and can be used for various studies related to interpretation and design of
such detectors.
\appendix
\section{Approximate factor for gain to scale 2D axisymmetric gain values to 3D values}
\label{sec:Appendix 1}

\textbf{Assumptions:}
The electric field can be assumed to have an average constant value throughout a GEM hole. The collection and extraction efficiency affect the gain in the same way for both the models.\\
\textbf{Additional information:}
Charges shared among the holes are known a priory.

For illustration, let us consider a double GEM configuration. From numerical simulation, it is
known that the charge sharing is $p_{c1}$ (central) : $p_{o1}$ (all others) for the top foil (towards the drift cathode), while it is $p_{c2}:p_{o2}$ for the bottom foil (towards the readout anode).
The gain in the central hole of the top layer is $G_{c1}$ that can be roughly estimated from the average field within the hole. It is the same for all other holes in a 3D model. So, if we assume 100
primaries collected by all the holes, the number of electrons coming out of the GEM holes in a 3D model should be 100 $G_{c1}$.

The gain in the top layer central hole of an axisymmetric model has the same gain, $G_{c1}$.
However, the gain in the other holes in the top layer is different since the electric field is
different there. From the field values, we can estimate that gain as, say, $G_{o1}$. Since the charge
sharing for the top layer is $p_{c1}:p_{o1}$, the number of electrons coming out of the top layer of the 2D
axisymmetric model is expected to be $p_{c1} G_{c1}$ + $p_{o1} G_{o1}$.

Let us denote the ratio between these two numbers, as $R_{1}$
\begin{equation}
\centering
 R_{1} = \frac{100 G_{c1}}{(p_{c1} G_{c1} + p_{o1} G_{o1} )} = \frac{100}{(p_{c1} + p_{o1} \times \frac{G_{o1}}{G_{c1}})}
\end{equation}
Following a similar chain of arguments and using the charge sharing data for the bottom GEM
foil, a similar ratio, $R_{2}$ , can be estimated.
\begin{equation}
\centering
R_{2} = \frac{100 G_{c2}}{(p_{c2} G_{c2} + p_{o2} G_{o2})} = \frac{100}{(p_{c2} + p_{o2} \times \frac{G_{o2}}{G_{c2}})}
\end{equation}
The scaling factor for estimating a 3D gain from 2D axisymmetric value thus given by,
\begin{equation}
\centering
 S_{G} = R_{1} \times R_{2}
\end{equation}
$G_{c}$-s and $G_{o}$-s have been estimated by finding an average field value within the GEM hole, and
assuming it to be a constant within the hole.
\acknowledgments
This work has been performed in the framework of RD51 collaboration. We wish to acknowledge the members of the RD51 Collaboration for their help and suggestions. We would like to acknowledge the necessary computing infrastructural help and support from SINP. We would also like to thank the respective funding agencies, DAE and INO collaboration. Author P.Bhattacharya acknowledges the University Grant Commission and Dr. D.S.Kothari Post Doctoral Scheme for the necessary support.


\end{document}